# TaiJi: State-Conditional Adaptive Combination of Machine-Learning and Physics-Based Global Weather Forecasts

**Jiale Wang[2], Yining Li[1], Guihua Wang[1,*]**

[1] **Department of Atmospheric and Oceanic Sciences & Institute of Atmospheric Sciences, Fudan University, Shanghai, China.**

[2] **School of Data Science, Fudan University, Shanghai, China.**

*Corresponding author: Guihua Wang (wanggh@fudan.edu.cn)

## Abstract

Machine-learning weather models now rival or surpass operational numerical weather prediction (NWP), yet no single model dominates across variables, pressure levels, lead times, or regions, and the returns from ever-larger individual models are diminishing. We present TaiJi (after the yin-yang concept of complementary opposites), a spatiotemporal adaptive ensemble framework that recasts optimal model-weight combination as a learned prediction problem: a lightweight convolutional neural network, conditioned on the recent atmospheric state, predicts at every grid point and lead time affine combination weights and an additive residual for the constituent forecasts. The combiner is trained end-to-end over Pangu-Weather, GraphCast, FuXi, IFS-HRES, and the IFS-ENS ensemble mean, a hybrid set of ML and NWP constituents, at less than a tenth of the cost of training any of its ML constituents. Under the WeatherBench 2 protocol (2020 test year), TaiJi outperforms the strongest baseline in nearly every combination of the eight core variables, lead times up to 10 days and two skill metrics (RMSE and ACC), and this advantage extends to heavy-precipitation extremes. The same state-conditional principle transfers to tropical-cyclone track forecasting: by fusing constituent tracks rather than gridded fields, TaiJi-TC reduces track error relative to the equally weighted multi-model consensus at lead times up to 96 h, by about 10% at 36 h, halves the slow along-track bias shared by all constituents up to 60 h, and lowers landfall position error by 13–15% at 60–84 h. The learned weights vary coherently with latitude, season and lead time rather than following a fixed rule. TaiJi thus offers an inexpensive route to better global forecasts that builds on, rather than replaces, existing forecast models.

# Introduction

Weather forecasting underpins activities as varied as agricultural planning, energy dispatch and the issuance of severe-weather warnings that protect lives and infrastructure[1]. For more than half a century, numerical weather prediction (NWP) has provided the scientific foundation of operational forecasting[2]. The European Centre for Medium-Range Weather Forecasts (ECMWF) Integrated Forecasting System (IFS), together with its ensemble counterpart IFS-ENS, encodes decades of progress in atmospheric dynamics, model physics and data assimilation, and remains the benchmark against which alternatives are judged[3,4]. Motivated in large part by the computational expense of operational NWP ensembles[5,6], data-driven forecasting has advanced rapidly in the past few years. FourCastNet[7] was among the first of a succession of high-resolution data-driven global models, whose prospects had long been debated[8] and which have since matched and, on many metrics, surpassed operational NWP[9]. Pangu-Weather[10] introduced hierarchical 3D Earth-specific attention[11–13] on pressure-level grids and outperformed the ECMWF high-resolution deterministic forecast (HRES) on standard upper-air metrics. GraphCast[14], a multi-mesh graph neural network[15,16] on icosahedral refinements, outperformed HRES on about 90% of 1380 verification targets. FengWu[17] and FuXi[18] extended ML forecast skill into the medium and extended range through multi-task learning and cascade architectures, and more recent models, namely NeuralGCM[19], Aurora[20] and GenCast[21], have extended the frontier to hybrid physics–ML solvers[22], atmospheric foundation models[23] and probabilistic generative forecasting. The resulting model population is methodologically diverse, differing in spatial discretization, temporal training strategy and learning objective, yet each system has been developed and evaluated largely in isolation.

Systematic evaluation reveals a consistent pattern. WeatherBench 2[24], the community standard for data-driven weather forecasting, shows that no single ML system leads uniformly across the full set of variables, pressure levels, forecast horizons and regions. Pangu-Weather has a distinct advantage in tropical-cyclone track forecasting, which has been attributed to its 3D Earth-Specific Transformer architecture and hierarchical temporal aggregation that together limit the error accumulation that typically degrades multi-day trajectory prediction[10,25]; GraphCast performs strongly on upper-air dynamical fields particularly at shorter lead times, where multi-mesh message passing[26] efficiently propagates synoptic-scale information; and IFS-ENS retains superior skill at extended lead times, in keeping with the well-documented role of ensemble averaging as a nonlinear filter that preserves predictable signal once deterministic skill saturates[4,27]. This suggests that the atmosphere, as a multi-scale, non-stationary and complex dynamical system[28,29], admits no single inductive bias[16] that confers a uniform advantage across the full global space–time domain; a bias that is advantageous in one region or at one scale typically carries a cost elsewhere.

Rather than designing yet another inductive bias and training a new forecasting model from scratch, one can therefore ask how the complementary knowledge already contained in existing models is best combined. The value of forecast combination is well established[30–32]. Within NWP, IFS-ENS and similar systems have long shown that calibrated multi-member forecasts outperform a single deterministic run, particularly at longer lead times[4,27,33]. Across operational centers, multi-model ensembles (MME) such as the TIGGE archive[34], the North American Multi-Model Ensemble (NMME)[35] and the Copernicus C3S multi-system seasonal service have repeatedly shown that combining structurally different forecasting systems is often more effective than refining any one of them, particularly in the medium-range, sub-seasonal and seasonal regimes. Statistical post-processing formalizes this idea under explicit objectives: deterministic superensembles[36,37] estimate bias-corrected linear weights from past performance, while Bayesian Model Averaging (BMA)[38] and Ensemble Model Output Statistics (EMOS)[39,40] cast the problem probabilistically, parametrizing a predictive distribution whose location and spread are linear functions of the ensemble members.

These classical schemes share an important limitation. The weights they assign are static: they are fixed in time and space once estimated, or at best allowed to vary slowly with lead time and location through independent regressions. None can recognize, for example, that GraphCast's multi-mesh representation is currently advantageous for a developing baroclinic wave over the North Atlantic while Pangu-Weather's local attention is better suited to the concurrent tropical convection over the Indian Ocean. As a result, a static combiner dilutes the contribution of the locally strongest model precisely when concentration would be most valuable. What is needed is a combiner that takes the current atmospheric state into account and allocates weights adaptively at every grid point and lead time.

A growing body of work has extended classical post-processing by applying neural networks to forecast calibration, with notable success for short- to medium-range NWP forecasts[41–44]. These methods, however, still rest on refining the output of a single forecasting system, rather than combining a heterogeneous pool of structurally different forecasting systems, and their accuracy is accordingly bounded by what that single source can deliver. The present work takes a different route: instead of refining one forecast, the combiner distributes weight among forecasts with different inductive biases. Prior MME work has shown such structural diversity to be valuable, and learned combination of data-driven weather models has recently begun to be explored[45,46]. In these approaches, however, the combination weights are inferred from the member forecasts themselves, together with lead time or recent errors, without access to the analyzed state of the atmosphere from which the forecasts evolve. A member's error, however, depends on the initial flow regime and not only on its output, so forecasts that look alike can differ widely in reliability. In information-theoretic terms, whenever the initial state $x_0$ carries information about the member errors e beyond that contained in the forecasts f, that is, $I(e; x_0 \mid f) > 0$, a combiner that sees only f can at best learn weights averaged over all flow regimes consistent with the same forecasts, and is therefore suboptimal (Methods). TaiJi instead performs state-conditional fusion: the background atmospheric state is supplied explicitly and modulates how the member forecasts are combined, over a hybrid pool of NWP and ML systems and across the full medium range.

We introduce TaiJi (named after the Chinese concept of *taiji*, the harmony of yin and yang, reflecting the aim of reconciling seemingly incompatible inductive biases), a spatiotemporal adaptive ensemble forecasting framework designed to address this gap. Its central premise rests on two ideas: first, that a suitably weighted combination of forecasts can outperform each of its members; and second, that the allocation of weights across constituent forecasts is itself a spatiotemporal prediction problem. We implement the TaiJi combiner as a lightweight dual-encoder U-Net[47]: one encoder maps the ERA5[48] atmospheric state at forecast initialization, and the other maps the forecasts of all constituents at all 20 lead times, into a shared high-dimensional feature space. A FiLM (Feature-wise Linear Modulation)[49] mechanism then uses the background-state embedding to modulate the deep feature maps of the constituent forecasts. The decoder outputs pixel-wise affine weight fields and an additive residual field, which are applied to the raw constituent forecasts to produce the final prediction. The combiner is trained end-to-end on the forecast objective, covering lead times from 12 to 240 hours at 12-hour intervals. In this study, it jointly spans two NWP systems (IFS-HRES and the IFS-ENS ensemble mean) and three state-of-the-art ML weather models (Pangu-Weather, GraphCast and FuXi). The TaiJi framework is agnostic to the specific constituent models: any atmospheric forecasting system can, in principle, be incorporated, provided its output conforms to a standardized data format.

Our contributions are threefold. First, we introduce a spatiotemporal adaptive ensemble framework that learns state-conditional combination weights, together with an additive residual correction, over a heterogeneous pool of operational NWP and ML weather models to yield an improved forecast. TaiJi's strongly regularized design allows the complete suite to be trained on two consumer-grade RTX 4090 GPUs in about 8 hours, at a training cost at least an order of magnitude below that of any of its ML constituents. Second, under the WeatherBench 2 protocol (2020 test year), TaiJi outperforms the strongest baseline on essentially every combination of the eight core evaluation variables, lead times of up to 10 days and two skill metrics (RMSE and ACC), and this advantage extends to heavy precipitation and, when the same principle is applied to the constituent tracks, to tropical-cyclone track and landfall forecasts. Third, we provide both an experimental and a theoretical analysis of pointwise affine combination. Experimentally, we find that the weight TaiJi assigns to each constituent varies with lead time and region: at short range it favors GraphCast; at longer lead times it favors IFS-ENS at high latitudes and FuXi in the tropics. Theoretically, we show a "no-harm" property: the optimal pointwise affine combiner has expected error no greater than that of any constituent at the same grid point and lead time, which provides a theoretical rationale for the design, although the advantage of the trained combiner must still be established empirically.

# Results

We evaluate TaiJi against five state-of-the-art baselines (Pangu-Weather, GraphCast, FuXi, IFS-HRES and the IFS-ENS ensemble mean) under the WeatherBench 2 protocol over the 2020 verification period. Forecast skill is quantified through three complementary metrics: root-mean-square error (RMSE) and anomaly correlation coefficient (ACC) for deterministic accuracy, and mean bias error (MBE) for the signed systematic deviation between forecast and verification. The evaluation spans eight representative variables: the 850 hPa zonal and meridional winds (U850, V850), 500 hPa geopotential (Z500), 700 hPa specific humidity (Q700), mean sea-level pressure (MSLP), 2 m temperature (T2m), and the 10 m zonal and meridional winds (U10, V10). These variables are widely used in operational forecasting and are of direct relevance to disaster prevention and mitigation, energy management, and agricultural planning. Forecast lead times range from 12 h to 10 days at 12 h intervals.

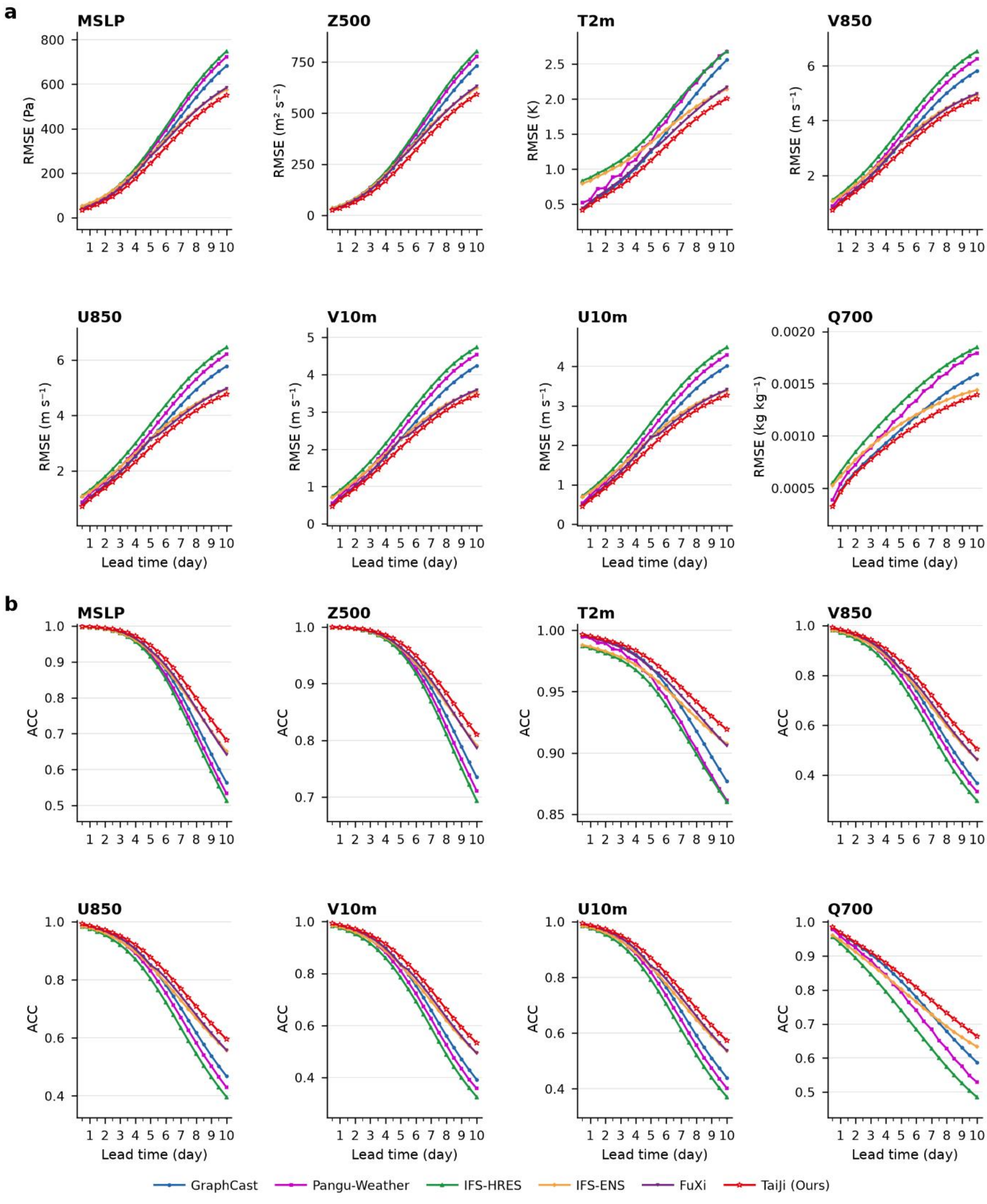

**Figure 1.** Deterministic forecast skill of TaiJi and five baselines under the WeatherBench 2 protocol. (a) Latitude-weighted global RMSE (lower is better) as a function of forecast lead time (12 h to 10 days) for each of the eight variables. (b) Latitude-weighted global ACC (higher is better) for the same variables. Each panel corresponds to one variable, and all results are computed over the 2020 verification period. FuXi does not forecast specific humidity and is therefore absent from the Q700 panels.

Figure 1 shows the principal result of this study: TaiJi attains the lowest RMSE and the highest ACC for all eight evaluated variables at essentially every lead time from 12 h to 10 days. Among the baselines, no single system maintains uniform leadership. GraphCast and Pangu-Weather lead at short lead times, plausibly because their architectures preserve initial-state accuracy during rollout: Pangu-Weather's three-dimensional Earth-specific transformer retains the vertical coupling across pressure levels while its hierarchical temporal aggregation limits autoregressive error accumulation, and GraphCast's multi-scale icosahedral graph avoids polar distortion and propagates information efficiently across scales. Beyond the short range, however, both fall behind, consistent with the tendency of iterative rollouts to compound one-step errors and of deterministic regression training to smooth the fields and progressively erode small-scale variability. FuXi, whose cascaded short-, medium- and long-range training strategy mitigates this error accumulation across temporal scales, is the strongest individual member in the medium range. The IFS-ENS mean regains its advantage at extended lead times. The IFS integrates the governing equations with physical parameterizations, which helps to maintain dynamical consistency over long integrations, and averaging over members that sample initial-condition and model uncertainty filters out the unpredictable components that dominate forecast error at these ranges. The ranking of the baselines also differs between variables.

TaiJi is designed for this situation, in which each baseline leads only in part of the variable–lead time space because no single inductive bias is complete. Instead of applying a fixed set of weights, TaiJi conditionally infers spatially resolved combination weights from the background atmospheric state and the predictions of every member. The weights are therefore re-derived for each forecast situation and shift continuously as the flow evolves, so that TaiJi can draw on whichever constituent is locally most skillful instead of averaging out the differences between models. In addition, a residual term learned on top of the weighted combination partly compensates for the smoothing that increases at long lead times. As a result, TaiJi improves on every baseline across essentially the entire variable–lead time space examined here. This is consistent with our hypothesis that an adaptive combination of existing models can reach a level of skill beyond that of any individual member.

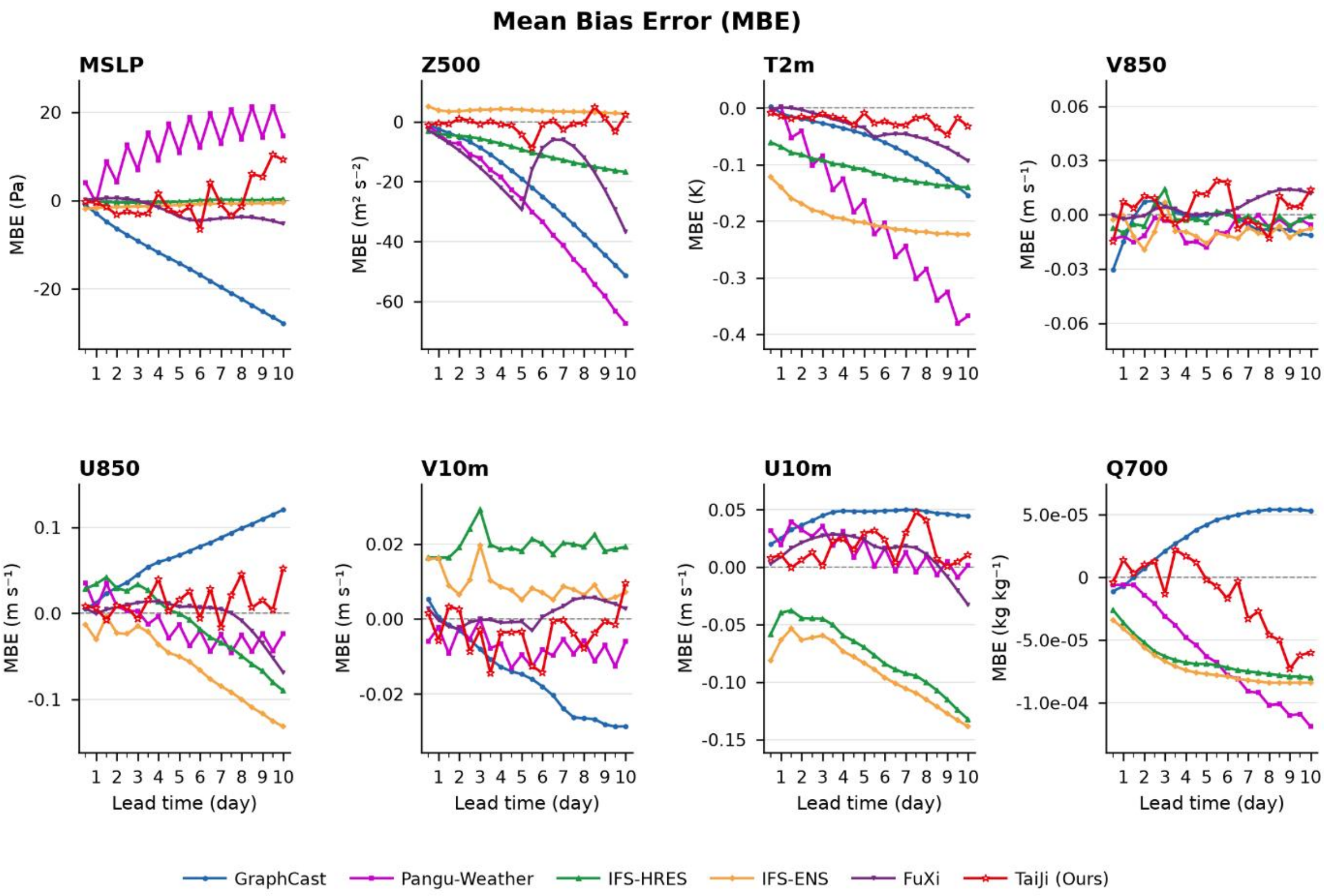

**Figure 2.** Mean bias error of TaiJi and the five constituent forecasts as a function of forecast lead time, shown for each of the eight evaluated variables. The horizontal zero-line denotes a perfectly unbiased forecast; positive values indicate systematic over-estimation, negative values systematic under-estimation. TaiJi (red) stays close to zero throughout the forecast horizon for seven of the eight variables.

Beyond RMSE and ACC, we assess systematic deviation using the mean bias error (MBE), defined as the signed, latitude-weighted global mean of the forecast residual (Figure 2). For seven of the eight variables, including Z500 and T2m, the TaiJi MBE stays close to the zero axis across all lead times, avoiding both the pronounced unidirectional drifts exhibited by several ML baselines and the low-frequency oscillations of the NWP members at long range. The only variable with a discernible offset is Q700, whose bias stays near zero at short-to-medium lead times but settles into a persistent negative bias at medium-to-long lead times; even so, its bias magnitude still compares favorably with that of the other models. The gains in RMSE and ACC are therefore not obtained at the cost of increased systematic bias. This is consistent with the complementary bias structures of the constituent models, whose signed errors partially cancel under adaptive weighting, and with the additive residual term that absorbs the small systematic component remaining after combination.

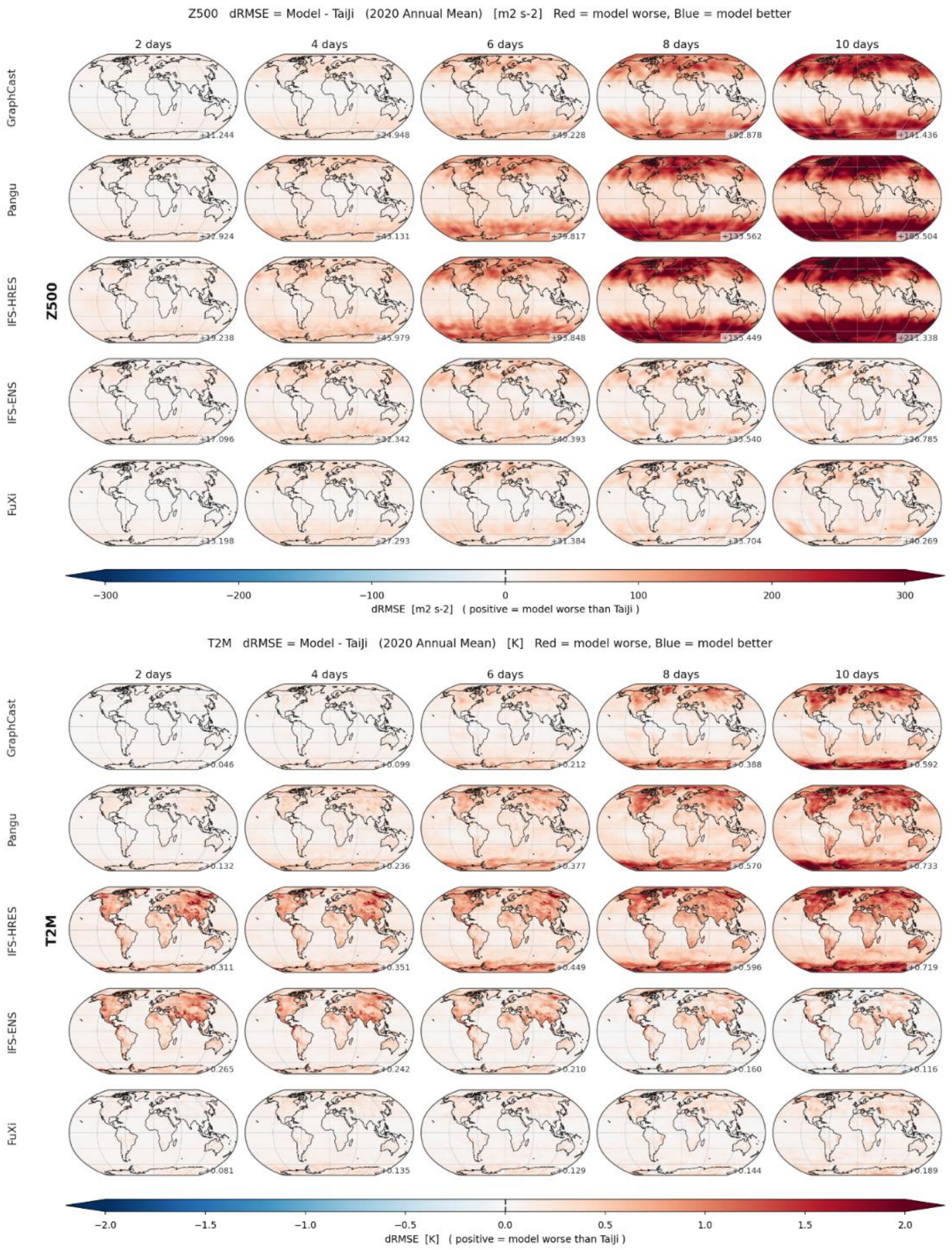

**Figure 3.** Spatial distribution of TaiJi's RMSE advantage over each comparison method, shown for Z500 (top) and T2m (bottom). Each map displays the difference (baseline RMSE − TaiJi RMSE) averaged across all forecast initializations in 2020, at lead times of 2, 4, 6, 8, and 10 days. Red shading indicates regions where TaiJi outperforms the corresponding baseline; blue would indicate the reverse. No discernible negative (blue) regions appear in any panel, indicating that TaiJi's advantage is spatially pervasive rather than confined to particular regions.

The latitude-weighted global means in Figure 1 show that TaiJi achieves the best aggregate skill, but a global average can mask regional deficits. Figure 3 shows that this is not the case. The annual-mean RMSE-difference fields are positive (red) at virtually every grid cell, for every comparison method, at every lead time examined, and for both Z500 and T2m: over 2020, TaiJi outperforms every baseline across virtually the entire globe, so the global advantage does not result from gains in some regions offsetting losses in others. Because these maps average over a full year of initializations, they reflect the typical skill at each location rather than the outcome of individual cases. That no baseline retains a region of superiority suggests that TaiJi adapts its combination to local conditions rather than merely selecting the best single model at each location, consistent with the no-harm property derived in Methods (Theoretical analysis). The margin is widest in the most demanding comparisons, namely against GraphCast and Pangu-Weather at long lead times and IFS-HRES in the medium range, where the individual skill of these models degrades and the gain from adaptive combination is greatest. For T2m, the large short-range margins over land relative to IFS-HRES and IFS-ENS should be read with the verification caveat noted in Methods: these systems are verified against ERA5 rather than against their own analysis, which inflates their short-range errors.

## Learned weights reveal physically coherent latitude- and lead-time-dependent specialization

A central question for any learned combination scheme is whether the inferred weights are physically meaningful or merely opportunistic fits to statistical regularities in the training data. We address this by analyzing the climatological structure of the TaiJi weight field along its two principal axes of variation: latitude and forecast lead time.

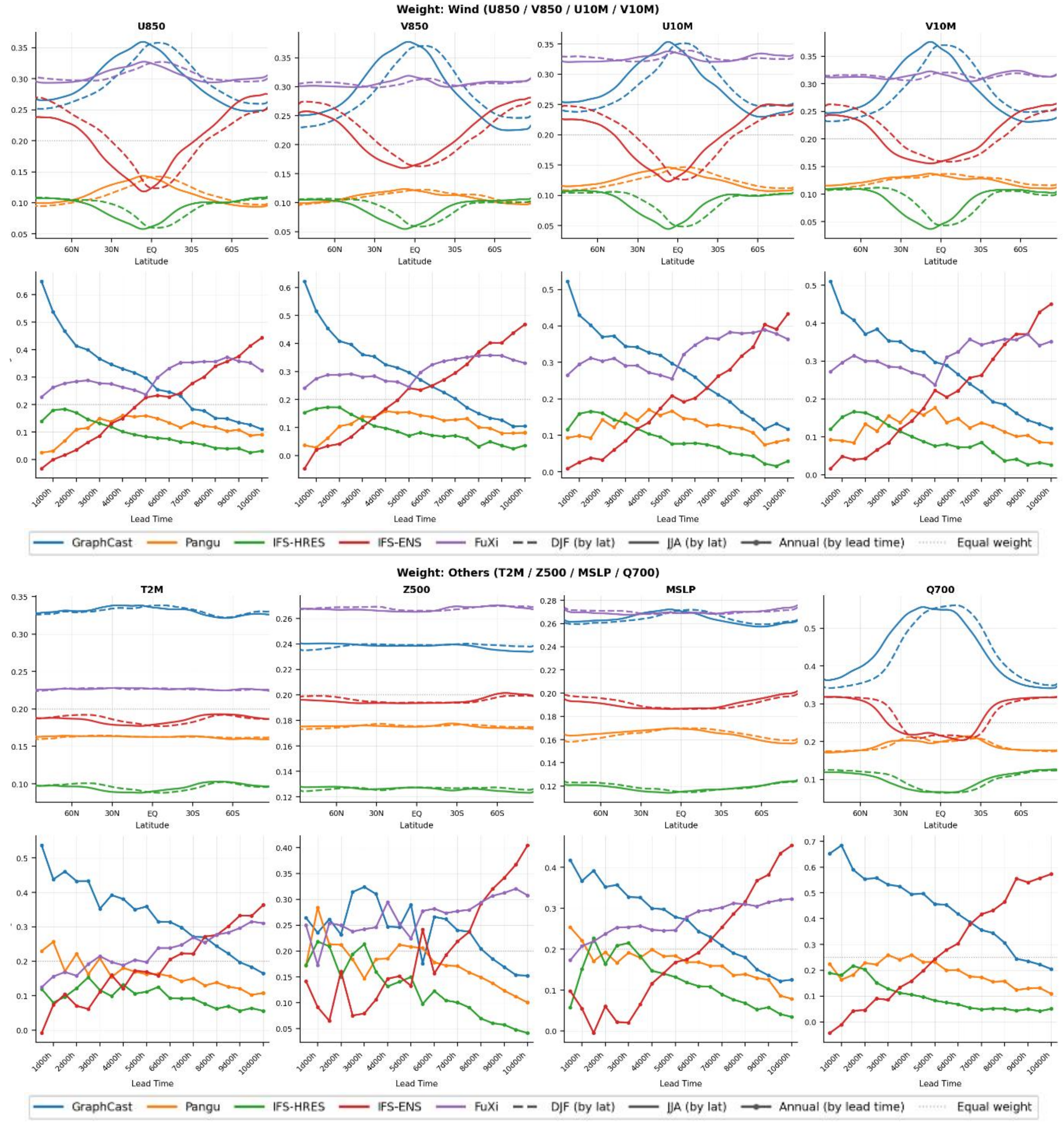

**Figure 4.** Climatological structure of TaiJi's combination weights for the eight evaluated variables. For each variable, the top sub-panel shows the mean assigned weight as a function of latitude, averaged across the verification period and the longitudinal dimension; solid curves correspond to boreal summer (JJA) and dashed curves to boreal winter (DJF). The bottom sub-panel shows the mean assigned weight as a function of forecast lead time, averaged across the verification period and all spatial dimensions. Curves are colored by constituent model.

Two main findings emerge from the latitudinal analysis (top sub-panels of Figure 4). First,

FuXi and GraphCast carry the highest mean weights overall, while IFS-HRES carries the lowest, a ranking that closely tracks the aggregate skill ordering established in Figure 1, indicating that a model's overall performance is reflected in the aggregate weight it receives within the TaiJi framework. Second, the weight allocations for the 10 m and 850 hPa wind fields exhibit a marked latitudinal dependence that runs in opposite directions for the ML and NWP members. ML methods receive systematically higher weights at low latitudes than at high latitudes, whereas the NWP members display the inverse pattern. A plausible physical interpretation is that high-latitude flow is dominated by quasi-linear, geostrophically balanced dynamics that are well represented by the dynamical core of the IFS, conferring an advantage on NWP in regimes that are closely governed by the resolved equations of motion. Tropical flow, by contrast, is shaped by strongly non-linear convective coupling and a high effective dimensionality that resists parametric closure, a regime in which the flexibility of data-driven models may be most valuable. That TaiJi recovers this division of labor without any prior instruction, learning its weight allocation purely from the forecast loss, suggests that the combiner extracts physically meaningful information about regime-dependent model competence directly from the raw atmospheric state.

Beyond the time-mean structure, the weights also migrate with the seasons. The contrast between the solid (JJA) and dashed (DJF) curves in Figure 4 shows that the latitudinal allocation described above shifts meridionally over the annual cycle. The signal is clearest in the wind variables and is sign-consistent wherever a hemispheric asymmetry is present: ML weights in the Northern Hemisphere are higher in JJA than in DJF (and lower in the Southern Hemisphere), while NWP weights show the opposite behavior. Thus, ML weights shift toward the summer hemisphere, whereas NWP weights shift toward the winter hemisphere. This mirrors the division of labor identified above, since the regimes each model class handles best, namely convectively driven non-linear dynamics for ML and quasi-balanced extratropical flow for NWP, move between hemispheres with the climatological solar forcing. TaiJi's seasonal weight shift tracks this migration, indicating that the combiner has learned the seasonal variability of model competence directly from the forecast loss. We regard this regime-tracking reading as the simplest interpretation consistent with the observed pattern, rather than as a causal attribution.

The lead-time-resolved weight curves in the bottom sub-panels of Figure 4 extend this interpretation along the temporal axis. GraphCast dominates the short-range regime, where its mean weight exceeds 50% and is well above that of any other constituent, before declining monotonically as the forecast horizon lengthens. IFS-ENS shows the opposite behavior: it carries the lowest weight at short range but rises monotonically with lead time to become the most heavily weighted member at day 10, consistent with the well-documented advantage of ensemble averaging at long range. FuXi receives relatively high, if irregular, weights at all lead times, in line with the aim of its cascaded short-, medium- and long-range training to maintain skill throughout the forecast range. Pangu-Weather contributes a moderate, stable weight concentrated in the medium range, while IFS-HRES rises from a small short-range contribution to a modest medium-range one before settling at low values at long lead times. Taken together, the constituents have complementary strengths that depend on lead time, which is the structure an adaptive combination is designed to exploit.

In addition to the weighting field, TaiJi appends a spatially resolved residual term to the weighted ensemble output, designed to correct systematic errors that survive the linear combination. Two aspects of its behavior support this design. First, the residual has a distinct spatial structure that is itself organized seasonally (top sub-panels of Figure 5). For the 850 hPa zonal wind, the residual amplitude at southern polar latitudes substantially exceeds that at northern polar latitudes in JJA, and the relationship reverses in DJF; the remaining variables show weaker but qualitatively similar shifts. This pattern is consistent with the residual canceling a seasonally varying component of the ensemble bias associated with the winter polar region, and it indicates that the residual acts as a state-conditional rather than a season-independent corrector. Second, the residual magnitude generally increases with lead time across the eight variables. This growth is consistent with the progressive smoothing of forecasts toward climatology at longer horizons, a bias that is prominent in the ML constituents and cannot be fully removed by reweighting alone. Because it is conditioned on the ERA5 state, the residual can restore part of the variance lost to this smoothing. Q700 is the only exception in sign: its residual becomes progressively more negative with lead time, reflecting a growing moisture over-estimation in the weighted forecast that the residual offsets; the slightly negative Q700 bias of TaiJi at long lead times (Figure 2) suggests that this correction becomes marginally too large. The residual term thus corrects structured biases that reweighting alone cannot remove.

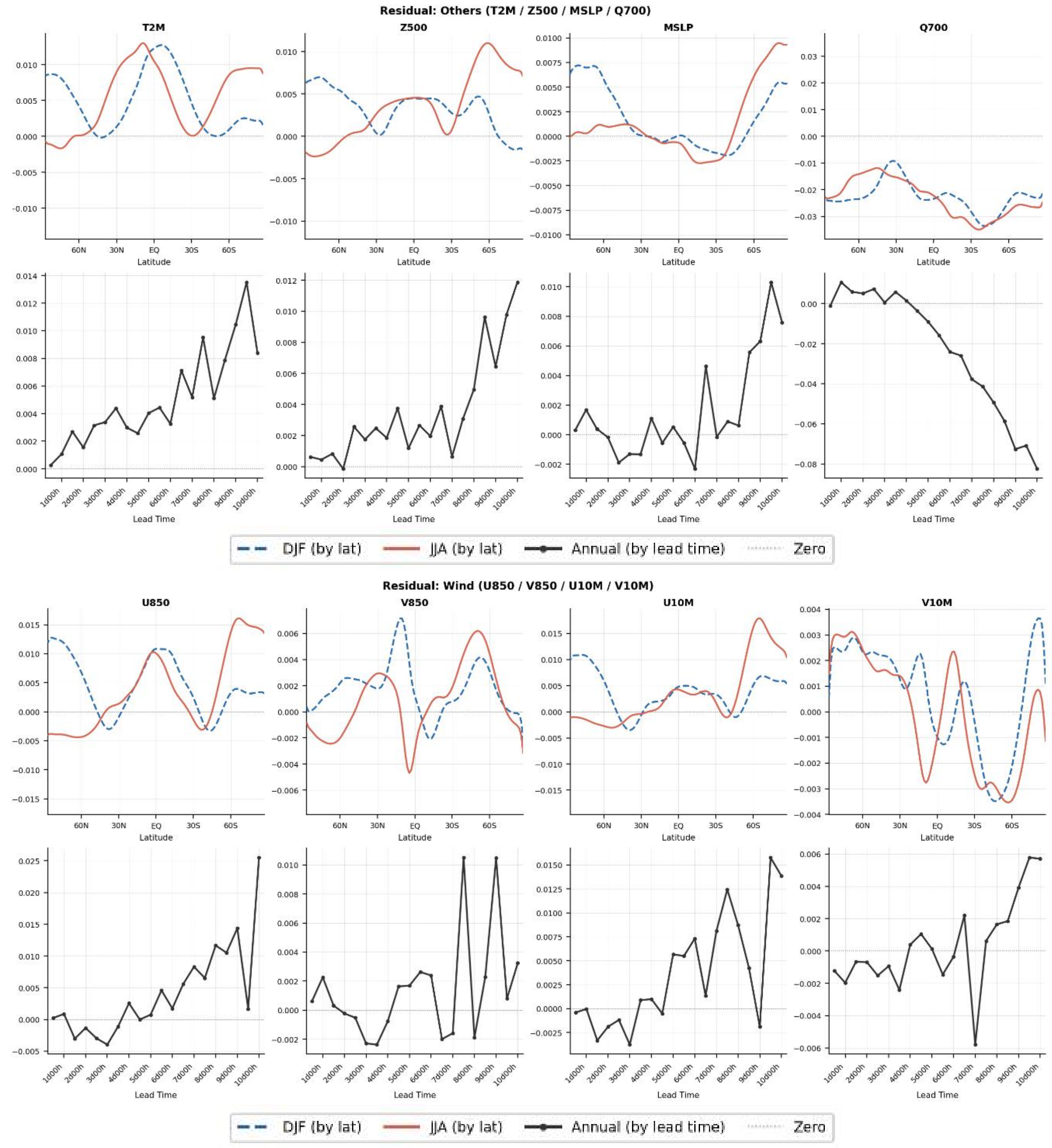

**Figure 5.** Climatological structure of TaiJi's learned residual correction for the eight evaluated variables. For each variable, the top sub-panel shows the longitude-averaged residual as a function of latitude, with solid curves for boreal summer (JJA) and dashed curves for boreal winter (DJF). The bottom sub-panel shows the residual magnitude as a function of forecast lead time, computed as the spatial root-mean-square of the additive correction field applied by TaiJi following the weighted ensemble combination, averaged across the verification period.

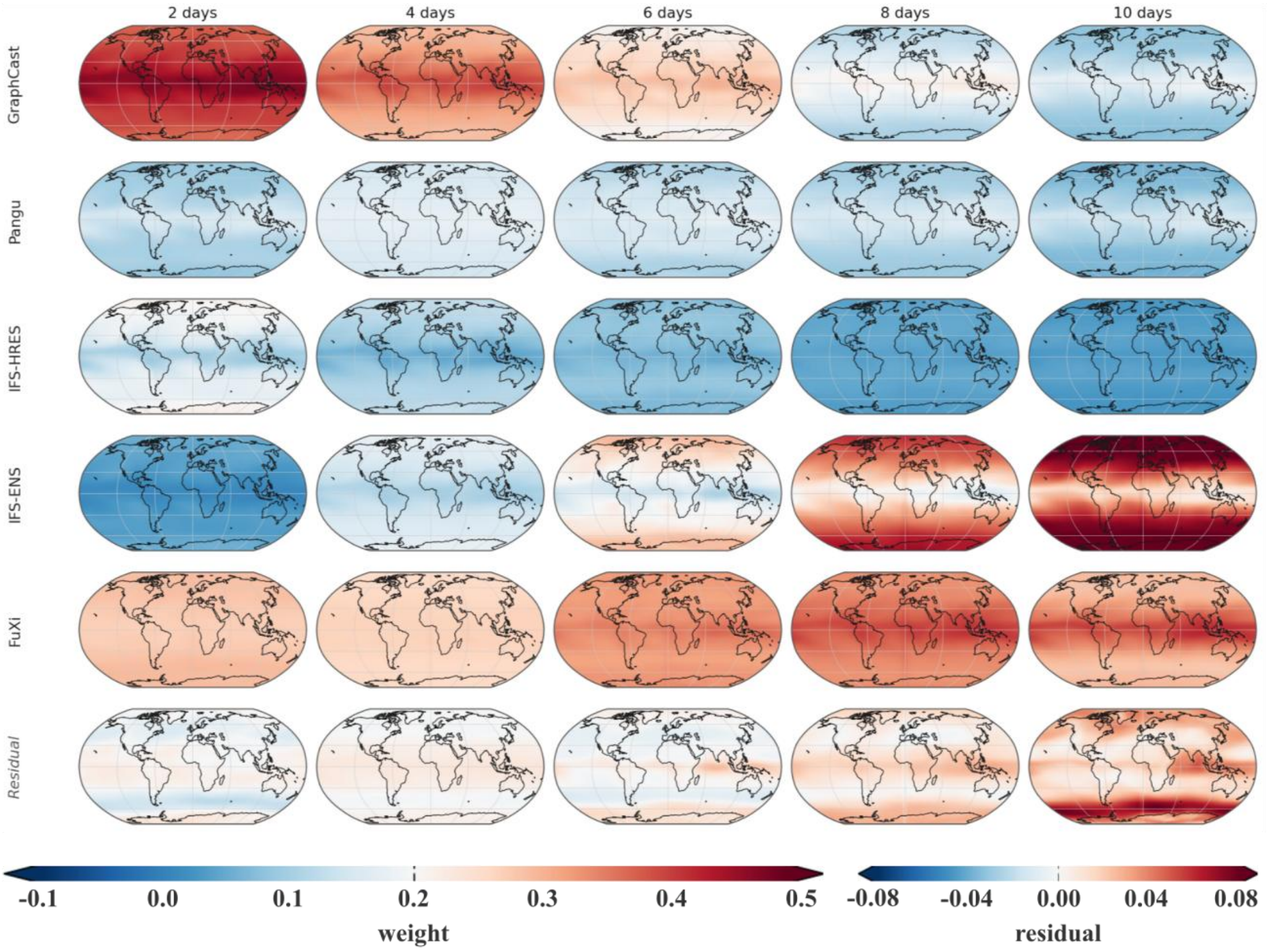

**Figure 6.** Pixel-wise weight and residual fields for the 850 hPa zonal wind (U850) across forecast lead times. The first five rows display the mean assigned weight for GraphCast, Pangu-Weather, IFS-HRES, IFS-ENS and FuXi, respectively; the bottom row displays the additive residual correction. Columns correspond to forecast lead times of 2, 4, 6, 8 and 10 days. All maps are averaged across the 2020 verification initializations. The weight color scale (left) and the residual color scale (right) are shown beneath the maps.

The latitudinal and lead-time decompositions presented above each project the four-dimensional combination structure onto a single axis, but the two dependences are not separable. Figure 6 illustrates this for U850 by displaying the full latitude–longitude weight field of every constituent at five selected lead times, together with the corresponding residual field. At the 2-day horizon the combiner assigns most weight to GraphCast and FuXi, but that weight is not uniformly distributed: GraphCast carries markedly higher weight at low latitudes than at high latitudes, while the FuXi contribution is more meridionally even. As the forecast horizon lengthens through the medium range the dominant pair shifts to FuXi and IFS-ENS, and the spatial pattern of the allocation changes accordingly: TaiJi assigns the bulk of the IFS-ENS contribution to the high-latitude regions where dynamical ensemble averaging is most informative, while reserving the bulk of the FuXi contribution for the low-latitude regions where FuXi retains most skill. The residual field is correspondingly state-dependent: its spatial structure migrates with lead time, growing in amplitude over the regions where the residual error of the weighted ensemble is largest at each horizon. The weighting is therefore neither a static blend nor a product of separate latitude and lead-time dependences; it varies jointly in space and lead time. This joint adaptation is likely to contribute to the skill advantage reported above, although isolating its contribution would require ablation experiments.

## Task-specific adaptation of TaiJi for high-impact weather

### TaiJi-Precip: fine-tuning for heavy precipitation

Precipitation is the most demanding test of tail behavior. The 6-h accumulated precipitation has a strongly heavy-tailed distribution (Fig. 7a): grid-point events of at least 30 mm $(6\text{ h})^{-1}$ make up only 0.06% of the samples, yet they contribute 3.1% of the total precipitation. Deterministic models trained with pointwise losses tend to produce smoothed fields and to underestimate heavy precipitation[50,51] because of the double-penalty effect[52,53]. Under an $L_1$ loss the optimal point forecast is the conditional median, which for a zero-inflated, heavy-tailed variable lies well below the conditional mean, so heavy events are expected to be underestimated even more strongly than under a squared-error loss. To address this, the first-stage combiner, trained with the latitude-weighted $L_1$ loss, is fine-tuned in a second stage. The second-stage loss, given in Equation (1), combines a soft CSI term, which rewards the detection of grid points exceeding the threshold, with a differentiable frequency-bias term, which penalizes departures of the predicted exceedance frequency from the observed one and so prevents the CSI term from being improved simply by over-forecasting.

$$\mathcal{L}_{\mathrm{CSI}} = 1 - \frac{\sum_j \sigma(\hat{y}_j - \tau)\,\sigma(y_j - \tau) + \varepsilon}{\sum_j [\sigma(\hat{y}_j - \tau) + \sigma(y_j - \tau) - \sigma(\hat{y}_j - \tau)\,\sigma(y_j - \tau)] + \varepsilon} + \lambda\left(\frac{\sum_j \sigma(\hat{y}_j - \tau) + \varepsilon}{\sum_j \sigma(y_j - \tau) + \varepsilon} - 1\right)^2 \quad (1)$$

where $\hat{y}_j$ and $y_j$ are the predicted and observed 6-h precipitation (mm) at grid point $j$, $\tau$ is the precipitation threshold ($\tau$ = 30 mm $(6\text{ h})^{-1}$ in this study), $\sigma(\cdot)$ is the logistic sigmoid, which replaces the non-differentiable indicator of exceeding $\tau$ with a smooth approximation, and $\varepsilon = 10^{-6}$ prevents division by zero. The second term is the squared deviation from unity of a soft frequency bias, defined as the ratio of the smoothed numbers of forecast and observed exceedances; because it uses the same sigmoid relaxation, the whole loss remains differentiable, and λ controls its weight relative to the CSI term. The loss is averaged over the samples in each batch.

Figure 7b, c compares the first-stage and fine-tuned TaiJi models with the constituents at the 30 mm (6 h)$^{-1}$ threshold, using the performance diagram (POD, SR and CSI), CSI as a function of lead time (12–84 h) and the frequency bias. Without any precipitation-specific training, the first-stage TaiJi already achieves a higher CSI than every constituent at 24–48 h and is comparable to the best constituent, GraphCast, at 12 h (Fig. 7c, left). The frequency bias shows that all constituents underestimate the frequency of such events, most strongly GraphCast, FuXi and IFS-ENS beyond 12 h (Fig. 7c, right). The first-stage TaiJi shares this tendency, and its frequency bias decreases with lead time, as expected for the conditional median favored by the $L_1$ loss. Second-stage fine-tuning with the loss of Equation (1) corrects this: from 24 h onward POD and CSI increase markedly, the two TaiJi variants perform comparably at 12 h, and the frequency bias of the fine-tuned model remains close to one at all lead times, so the higher CSI is not obtained by over-forecasting (Fig. 7b, c). The 48 h forecast valid at 00 UTC 17 December 2020, when Tropical Cyclone Yasa was approaching Fiji, illustrates the effect (Fig. 7d, e): the fine-tuned TaiJi reproduces the location and intensity of the rainfall cores in the South Pacific more closely than the constituent models, most of which produce weaker and more diffuse maxima.

These results have two implications. First, without any precipitation-specific training, the adaptive combination already matches or exceeds the best constituent at every lead time. Second, because this skill resides in a small trainable combiner, it can be specialized at very low cost by a second fine-tuning stage, which here substantially improves the forecast of heavy precipitation beyond the shortest lead time. The same specialization is much harder to obtain from the constituents themselves: fine-tuning a large AI weather model toward a narrow objective is computationally expensive and risks degrading its general skill or collapsing its output distribution, and operational NWP systems offer essentially no practical route to objective-specific retuning. TaiJi avoids both difficulties and provides an inexpensive two-stage workflow, in which a general-purpose combination is learned first and then adapted to a particular hazard, region or operational requirement.

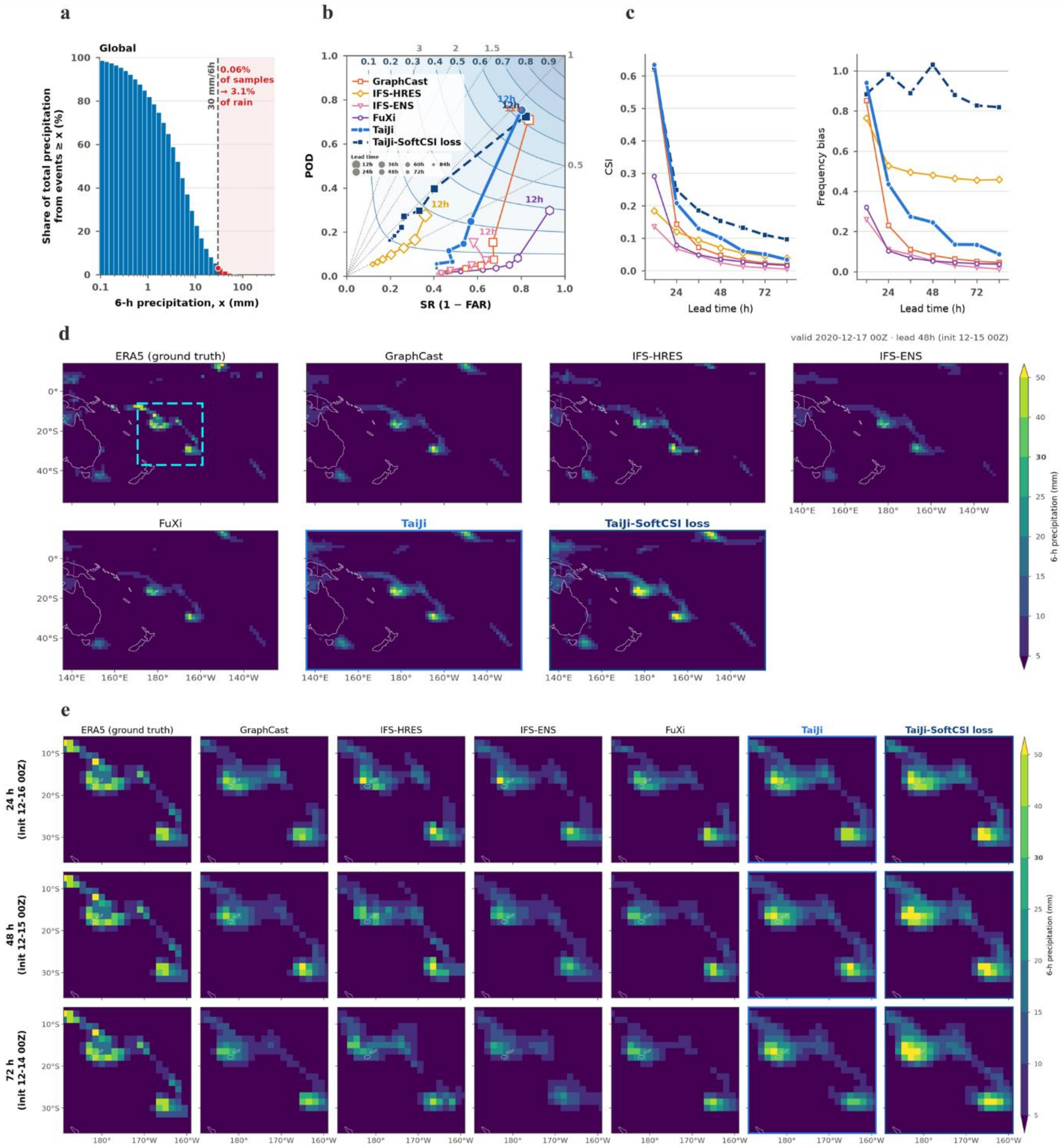

**Figure 7.** Verification of global 6-h accumulated precipitation for all forecast initializations in 2020. TaiJi denotes the first-stage combiner trained with the latitude-weighted L1 loss, and TaiJi-SoftCSI loss the model after second-stage fine-tuning with the loss of Equation (1); Pangu-Weather does not forecast precipitation and is therefore not included. (a) Share of total precipitation contributed by events at or above a given 6-h accumulation; events of at least 30 mm (6 h)$^{-1}$ make up 0.06% of the samples but contribute 3.1% of the precipitation. (b) Performance diagram of the probability of detection (POD), success ratio (SR) and critical success index (CSI) at the 30 mm (6 h)$^{-1}$ threshold for lead times of 12–84 h. (c) CSI (left) and frequency bias (right) at the same threshold as functions of lead time. (d) 6-h

accumulated precipitation over the South Pacific valid at 00 UTC 17 December 2020, 48 h after initialization; the dashed box in the ERA5 panel marks the region enlarged in (e). (e) Enlarged view of the boxed region at the same valid time for lead times of 24, 48 and 72 h (initialized at 00 UTC on 16, 15 and 14 December 2020, respectively).

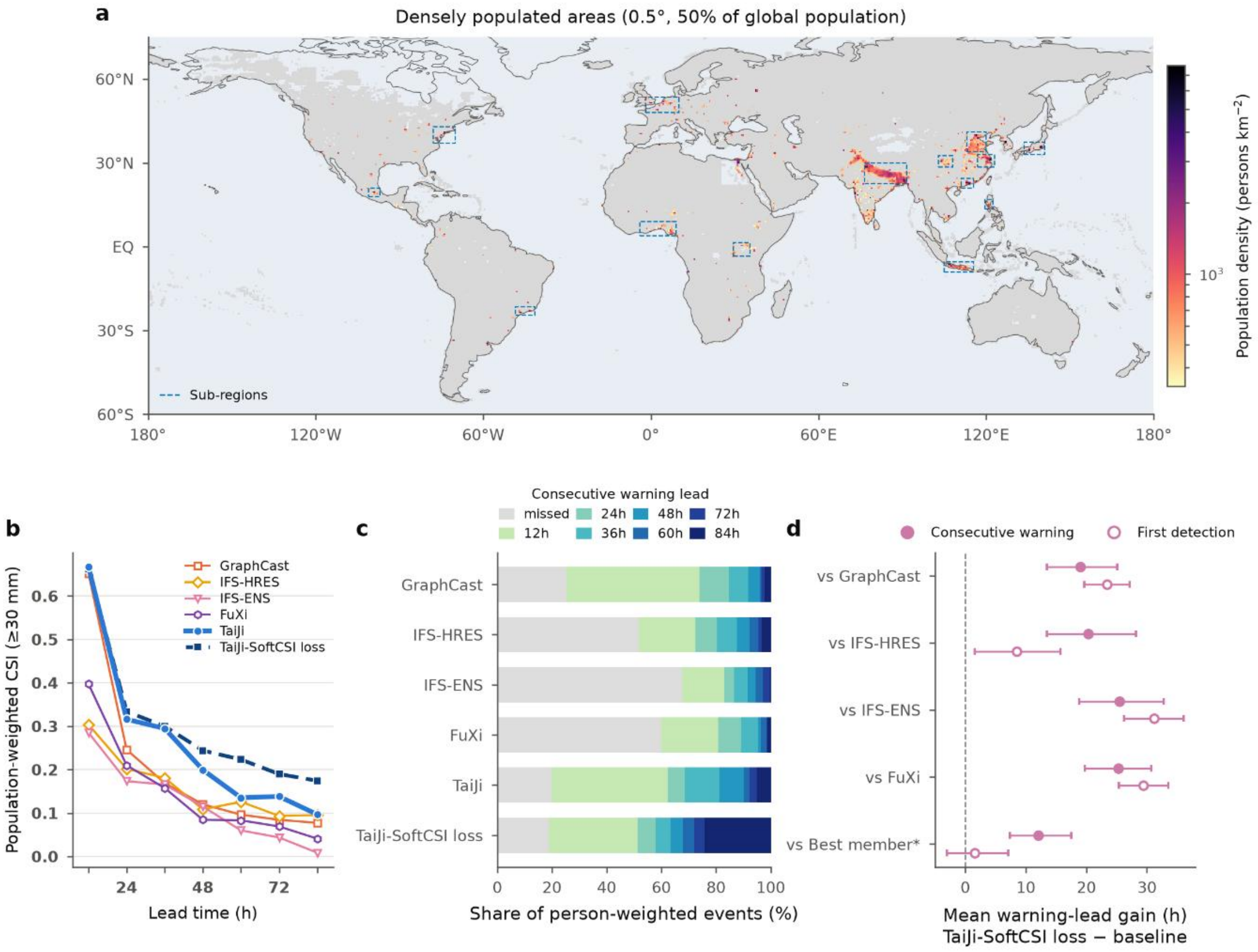

**Figure 8.** Population-weighted verification of heavy-precipitation forecasts over densely populated areas worldwide, against ERA5 for 2020. (a) Densely populated areas, defined as the 0.5° cells that together contain 50% of the global population (GPWv4.11); dashed boxes mark the sub-regions. (b) Population-weighted CSI at 30 mm (6 h)$^{-1}$. (c) Distribution of the consecutive warning lead time, defined as the longest lead time up to which all forecasts from 12 h onward exceed the threshold. (d) Mean gain of the fine-tuned TaiJi (TaiJi-SoftCSI loss) in consecutive (filled) and first-detection (open) warning lead time. *Best member: for each event, the constituent with the longest lead time, selected in hindsight. Shading and error bars denote 95% block-bootstrap confidence intervals.

The verification above weights all grid points equally, whereas the potential consequences of heavy precipitation depend on where it occurs. As a complementary diagnostic, we therefore repeated the evaluation over densely populated areas, defined as the 0.5° cells that together contain half of the world's population in the GPWv4.11 dataset[54] (Fig. 8a). Each 1.5° grid cell is weighted by the population of its densely populated sub-cells. Under this weighting, both TaiJi variants match or exceed every constituent in CSI from 24 h onward, the fine-tuned TaiJi has the highest CSI from 48 h onward, and at 12 h it performs comparably to GraphCast and the first-stage TaiJi (Fig. 8b). The distribution of consecutive warning lead times is consistent with these differences (Fig. 8c): both TaiJi variants leave a smaller share of population-weighted events undetected than any constituent, and the fine-tuned TaiJi has the largest share of events for which the exceedance is forecast continuously from 84 h onward.

A paired, event-by-event comparison supports these results (Fig. 8d; 95% confidence intervals from a block bootstrap[55]). For consecutive warnings, the mean lead-time gain of the fine-tuned TaiJi is positive relative to every individual constituent, and the corresponding confidence intervals exclude zero. For first detection, the gain is also positive relative to every constituent, although it is smaller relative to IFS-HRES than relative to the other three. We further compared the fine-tuned TaiJi with a hindsight benchmark that selects, for each event, the constituent with the earliest warning; because this selection is not possible in real time, the benchmark is an optimistic reference. Against it, the gain remains positive for consecutive warnings, whereas the difference in first detection is not statistically significant. Relative to the best constituent of each case, the advantage of the fine-tuned TaiJi therefore lies mainly in forecasting the exceedance consistently across lead times rather than in earlier first detection. These results are based on a single year of events, verified against ERA5 at 1.5° resolution and weighted by gridded population estimates. They describe the population exposed to correctly forecast heavy precipitation and should not be interpreted as estimates of avoided impacts or losses.

## TaiJi-TC: adapting TaiJi for tropical-cyclone track forecasting

Cyclone tracks expose a limitation of combining fields. On the 1.5° grid of TaiJi[24], the inner core of a cyclone spans only one or two grid cells, so a tracker applied to the combined fields would add position errors comparable to the differences among the constituents. TaiJi-TC therefore fuses the constituent tracks themselves, extracted from 0.25° forecast fields[56], while keeping the principle of TaiJi: member weights conditioned on the initial atmospheric state, plus a correction for errors that all members share. Because a season offers only a few hundred storm-centered forecasts, track fusion is learned jointly with the fusion of six storm-environment fields under ERA5 supervision, following recent cyclone models trained on reanalysis and best tracks together[57]. Forecasts are verified against IBTrACS[58] and compared with the equally weighted consensus of the five constituents, the standard benchmark for multi-model track guidance[59].

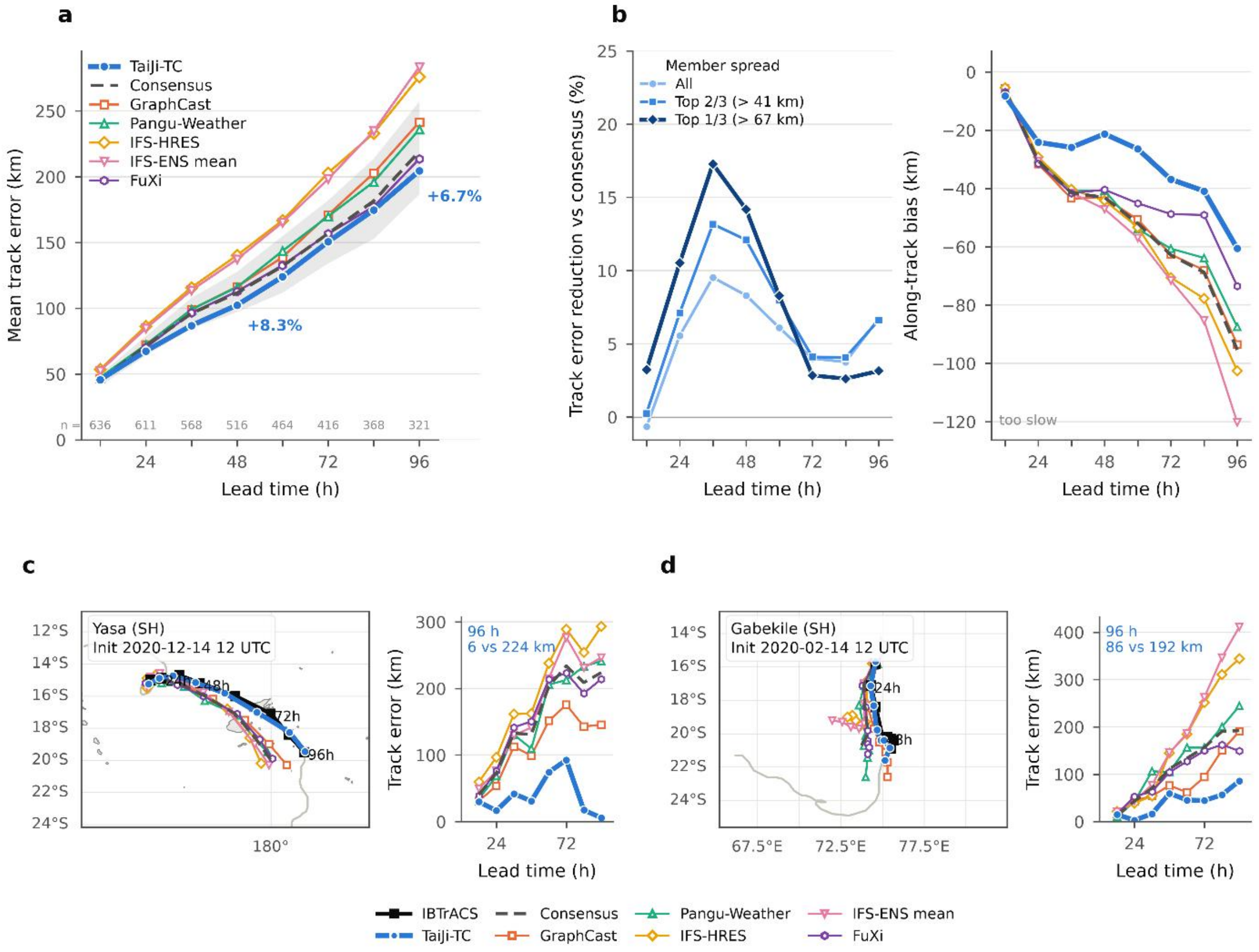

**Figure 9.** Tropical-cyclone track forecasts for 2020. (a) Mean track error versus lead time on a homogeneous sample; numbers above the axis give the sample size, and percentages give the error reduction of TaiJi-TC relative to the consensus. (b) Left: error reduction relative to the consensus for all cases and for the upper two-thirds and upper third of member spread

(mean distance of members from the consensus). Right: mean along-track bias (negative, too slow); quasi-stationary cases are excluded. (c, d) Example forecasts for Yasa and Gabekile: tracks (left) and track error (right). Shading denotes 95% storm-level bootstrap confidence intervals.

TaiJi-TC has the lowest track error of all methods from 24 h onward and matches the consensus at 12 h (Fig. 9a). Its advantage over the consensus peaks near 10% at 36 h and persists to 96 h (8.3% at 48 h, 6.7% at 96 h), whereas by 96 h the consensus is no better than its best member, FuXi. The gain is largest exactly where equal weighting is least justified. For the third of cases in which the members disagree most, the reduction reaches 17% at 36 h, about twice the sample mean (Fig. 9b, left). Consensus error is known to grow with member spread[60]; TaiJi-TC turns this disagreement into skill by using the initial state to decide which members to trust. Beyond 60 h the conditional advantage fades, as spread increasingly reflects uncertainty that the initial state cannot resolve.

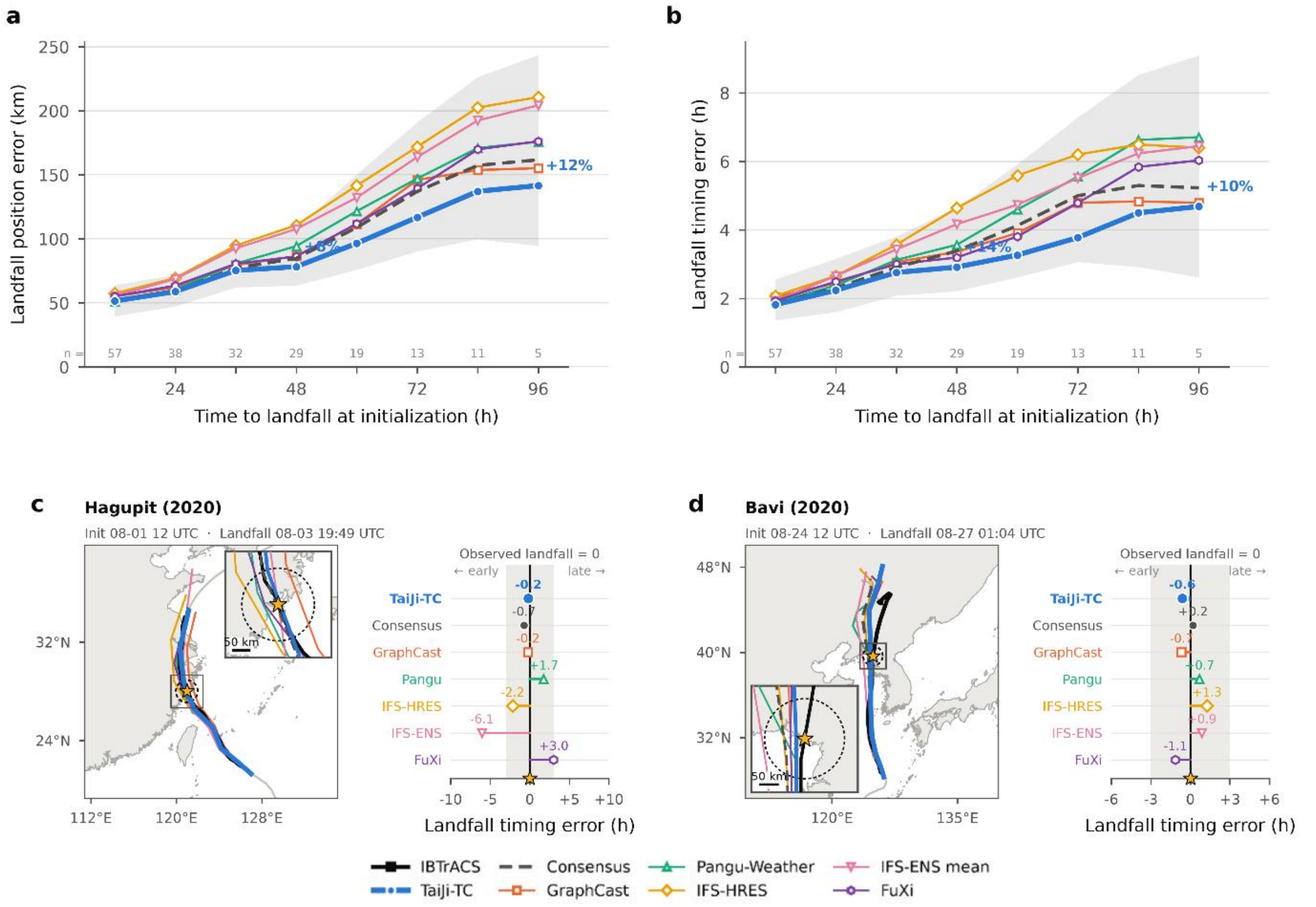

**Figure 10.** Landfall forecasts for 2020. (a) Landfall position error and (b) landfall timing error versus time to landfall at initialization, evaluated at the observed landfall time; shading denotes 95% storm-level bootstrap confidence intervals, and percentages give the error

reduction of TaiJi-TC relative to the consensus. (c, d) Example forecasts for Hagupit and Bavi: tracks, with the observed landfall point (star) and a 100-km radius (left), and landfall timing errors (right; gray shading, ±3 h).

The along-track decomposition reveals the most consistent source of the gain (Fig. 9b, right). All five constituents move cyclones too slowly, with an along-track bias that grows to between −73 and −120 km at 96 h. Because every member shares this bias, no weighted average of their positions can remove it. TaiJi-TC halves it up to 60 h (−21 km against −42 km for the consensus at 48 h) through its unnormalized weights and learned displacement, the track counterpart of the residual correction in TaiJi. Yasa and Gabekile show the effect in individual forecasts (Fig. 9c, d): while the members lag behind or diverge from the observed storm, with 96-h errors of 150–410 km, TaiJi-TC stays within 100 km of the best track throughout.

The track gains carry over to landfall, where errors matter most. Rather than locating where each forecast first crosses the coast[61], which leaves forecasts that stall or recurve offshore unverified, we evaluate all methods at the observed landfall time. TaiJi-TC has the smallest landfall position error at every lead time, 13–15% below the consensus at 60–84 h (Fig. 10a), and its timing error falls even more, by 20–24% at 60–72 h (Fig. 10b), as expected from the correction of the slow bias; beyond 72 h the samples are small and the gains indicative. Hagupit and Bavi illustrate two situations behind these numbers (Fig. 10c, d). For Hagupit, early and late member errors cancel in the consensus (14 km), and TaiJi-TC halves what remains (7 km). For Bavi, all members pass west of the landfall point, so averaging cannot help (51 km), yet TaiJi-TC moves the forecast toward the observed landfall (37 km).

## Ablation studies

To isolate the contribution of each design choice to TaiJi's overall performance, we conduct three systematic ablations: on the combination strategy, on the value range permitted for the constituent weights, and on the residual correction module. The complete results, with RMSE and ACC for all eight variables averaged over the 20 lead times from 12 to 240 h, are reported in Tables 1 and 2 (wind variables and the remaining variables, respectively).

**Table 1.** Ablation results for the wind variables (V850, U850, V10 and U10). RMSE (m $s^{-1}$) and ACC, each averaged over all 20 lead times (12–240 h), are reported for each ablation group: combination strategy (correct-then-combine versus combine-then-correct), weight value range (four constraint regimes) and inclusion of the residual correction term. Bold marks the best result within each group (lower RMSE, higher ACC). In the tables, "ensemble" refers to the weighted combination and "bias" to the residual term.

| Group | Setting | V850 | | U850 | | V10 | | U10 | |
|---|---|---|---|---|---|---|---|---|---|
| | | RMSE | ACC | RMSE | ACC | RMSE | ACC | RMSE | ACC |
| Ensemble | Ensemble-then-correct | 2.92 | 0.801 | 2.88 | 0.836 | 2.07 | 0.814 | 1.98 | 0.827 |
| | Correct-then-ensemble | **2.87** | **0.805** | **2.83** | **0.842** | **2.01** | **0.820** | **1.95** | **0.831** |
| Weight | $(-\infty, +\infty)$ | **2.92** | **0.801** | **2.88** | **0.836** | **2.07** | **0.814** | **1.98** | **0.827** |
| | $[-1, 1]$ | 2.95 | 0.799 | 2.90 | 0.834 | 2.09 | 0.811 | 2.00 | 0.825 |
| | $(0, +\infty)$ | 3.02 | 0.793 | 2.96 | 0.829 | 2.13 | 0.807 | 2.04 | 0.819 |
| | $[0, 1]$ | 3.06 | 0.789 | 3.03 | 0.823 | 2.17 | 0.802 | 2.08 | 0.814 |
| Bias | With bias | **2.92** | **0.801** | **2.88** | **0.836** | **2.07** | **0.814** | **1.98** | **0.827** |
| | Without bias | 3.12 | 0.784 | 3.09 | 0.818 | 2.21 | 0.797 | 2.11 | 0.810 |

**Table 2.** Ablation results for MSLP, Z500, T2m and Q700, following the conventions of Table 1. RMSE is given in Pa for MSLP, $m^2$ $s^{-2}$ for Z500, K for T2m and g $kg^{-1}$ for Q700.

| Group | Setting | MSLP | | Z500 | | T2M | | Q700 | |
|---|---|---|---|---|---|---|---|---|---|
| | | RMSE | ACC | RMSE | ACC | RMSE | ACC | RMSE | ACC |
| Ensemble | Ensemble-then-correct | 273 | 0.897 | 279.9 | 0.941 | 1.20 | 0.967 | 0.98 | 0.830 |
| | Correct-then-ensemble | **270** | **0.901** | **278.4** | **0.944** | **1.18** | **0.971** | **0.96** | **0.834** |
| Weight | $(-\infty, +\infty)$ | **273** | **0.897** | **279.9** | **0.941** | **1.20** | **0.967** | **0.98** | **0.830** |
| | $[-1, 1]$ | 275 | 0.895 | 283.3 | 0.938 | 1.21 | 0.964 | 0.99 | 0.827 |
| | $(0, +\infty)$ | 282 | 0.888 | 287.5 | 0.933 | 1.24 | 0.958 | 1.01 | 0.823 |
| | $[0, 1]$ | 287 | 0.883 | 293.3 | 0.927 | 1.26 | 0.952 | 1.03 | 0.817 |
| Bias | With bias | **273** | **0.897** | **279.9** | **0.941** | **1.20** | **0.967** | **0.98** | **0.830** |
| | Without bias | 293 | 0.877 | 297.8 | 0.923 | 1.29 | 0.946 | 1.05 | 0.812 |

The first ablation examines the combination strategy. The TaiJi framework as presented adopts a combine-then-correct paradigm, in which the weighted ensemble output is computed first and the residual correction applied subsequently. This ordering was chosen because it factorizes the combiner into a linear weighting stage and a separate correction stage, which renders the role of each module transparent, allows the weights to be read directly as the trust placed in each raw constituent, and supports rapid convergence under the $L_1$ objective. We compare this configuration against the inverse correct-then-combine arrangement, in which a learned bias correction is applied to each member individually before the weighted ensemble operates on the bias-corrected predictions. The correct-then-combine arrangement yields a small but consistent reduction in RMSE for every variable, suggesting that the individual error structures of the constituent models are sufficiently distinct that pre-conditioning each forecast to a common reference benefits the subsequent linear combination. Its ACC, however, is slightly lower than that of combine-then-correct for every variable, so the gain is confined to error magnitude and does not extend to anomaly correlation. We nevertheless retain combine-then-correct as the main configuration. Under correct-then-combine the weights act on bias-corrected rather than raw members, so they no longer measure the trust placed in each constituent, and the member-wise corrections enter the forecast only through their weighted sum, so they cannot be attributed to individual members. Because the weight and residual diagnostics in Figures 4–6 are a central part of this study, we accept a small increase in RMSE in exchange for a directly interpretable combiner; the principal claims of the paper therefore rest on the simpler configuration. We regard the correct-then-combine ordering as the most promising refinement of the architecture for future work.

The second ablation examines the value range permitted for the constituent weights. Four constraints are evaluated: $w \in [0, 1]$ (a convex combination, the classical post-processing constraint), $w \in [-1, 1]$ (signed but bounded), $w \in (0, +\infty)$ (positive but unbounded), and $w \in (-\infty, +\infty)$ (the configuration adopted in this work). The same skill ordering holds for every variable in Tables 1 and 2: $(-\infty, +\infty) > [-1, 1] > (0, +\infty) > [0, 1]$. Both transitions in this ordering can be interpreted. The first, namely that admitting negative weights improves the ensemble, is consistent with the deterministic superensemble literature[36,37], in which regression weights are not restricted to the simplex, and contrasts with mixture-based schemes such as Bayesian Model Averaging[38], whose weights must form a convex combination for the predictive mixture to remain a valid probability density. From an optimization standpoint, the convex constraint reduces the feasible weight set to a low-dimensional simplex that excludes precisely those configurations required to cancel correlated biases between members; permitting $w < 0$ allows the combiner to offset members whose errors are correlated with those of others in a given regime, moving the solution toward the unconstrained least-squares optimum. A physical reading is also possible: when two constituent models share a coherent error of the same sign in a particular regime (for example, a systematic over-prediction of tropical convective intensity), a negative weight on one of them allows the shared component to be partly canceled against members that do not carry it, which is impossible under a convex constraint.

The second transition, namely that lifting the upper bound from unity to infinity yields further improvement, has an analogous explanation. The convex constraint $w \leq 1$ implicitly assumes that the truth lies within the convex hull of the ensemble members. When this assumption fails (as it does, for instance, when every member systematically under-resolves a sharp feature), the optimal estimate of the truth requires amplifying one or more members beyond unit weight, an operation that the unbounded configuration permits and the convex one forbids. That weights above unity are selected by the learned combiner suggests that, in some regimes, the verification field lies outside the convex hull of the constituent forecasts; because weights above unity can also arise from compensating combinations of correlated members, this reading should be regarded as indicative rather than conclusive.

The third ablation isolates the contribution of the residual correction module by training a variant in which the residual is set identically to zero, leaving only the weighted ensemble combination in operation. Removing the residual produces a substantial degradation in performance across every variable. This indicates that, notwithstanding the complementary error structures of the five constituents, there remains a common component of the forecast error (a systematic under-estimation or coherent mis-positioning that survives any linear combination of the available members) that requires a state-dependent correction. The residual term provides this correction, conditioned on the ERA5 state; without it, TaiJi has no means of removing the bias shared by all constituents. The weighting and correction modules are therefore complementary: the former exploits the differences among the constituents, and the latter corrects the errors they have in common.

# Discussion

## Combining inductive biases with a lightweight learned combiner

TaiJi is motivated by a property of the current population of forecast models rather than by a deficiency of any individual model. Each operational forecast system encodes a distinct inductive bias about atmospheric evolution: Pangu-Weather through a three-dimensional Earth-specific attention geometry, GraphCast through multi-mesh message passing, FuXi through cascaded specialization by lead time, and the IFS through the discretized primitive equations and their parameterizations. These are independently derived hypotheses about the structure of the atmosphere, and Figures 1 and 4 show that none dominates uniformly; leadership migrates across variables, lead times, regions and seasons. Such a population contains predictive information that no single member provides, and learning how to combine the existing members is a direct way of exploiting it.

The approach is deliberately asymmetric. Constituent parameters, ranging from tens of millions to billions of weights for the ML models, are held fixed and never enter the optimization. The only component that is trained is the combiner, a U-shaped network with an asymmetric dual encoder whose training requires only a small fraction of the compute used to train any of the ML constituents. Its output is restricted to a small, directly interpretable set of combination coefficients and a residual at each grid point, defined on top of fixed, high-dimensional models, and the results above show that this suffices for the combination to outperform every one of its constituents.

This design also makes the best skill attainable from existing models inexpensive to reach and easy to build on. With constituent forecasts taken from public archives, a system that surpasses the best available guidance can be trained on modest hardware. Because this skill resides in a small trainable module, downstream tasks can be attached to it directly: task-specific objectives, regional refinement and extreme-event specialization can each be addressed by fine-tuning the combiner, starting from state-of-the-art forecast quality without access to, or retraining of, any of the underlying models.

## Per-task decomposition as strong regularization

TaiJi trains a separate combiner for each of the eight target variables rather than a single network producing the entire output volume, a choice that departs from common practice.

The eight variables differ by orders of magnitude in physical units and dynamical range

(geopotential in $m^2$ $s^{-2}$, temperature in K, specific humidity in kg $kg^{-1}$, winds in m $s^{-1}$, pressure in Pa). A network producing them jointly must reconcile loss surfaces whose gradients are not aligned, which in practice is handled by hand-tuned per-variable loss weights. This tuning is never fully resolved, because the objectives conflict and any fixed weighting improves one variable at the expense of another; a substantial part of the engineering effort in training large forecast models is absorbed by the search for a weighting under which no variable is unacceptably degraded. Monolithic models accept this trade-off because training one large model per variable would be prohibitive. For a combiner with negligible per-task cost the trade-off is unnecessary, and decomposing the problem removes inter-variable competition by construction. The eight tasks are also mutually independent and trained in parallel, so the full suite can be trained on modest hardware.

The decomposition further acts as a strong regularizer, and this is its principal benefit. Each combiner addresses a single narrowly specified regression problem whose hypothesis space is anchored to the neighborhood of the constituent forecasts, so the solution is determined largely by that constraint rather than by the volume of training data. Sample complexity is correspondingly low[62,63], which allows TaiJi to be trained in a deliberately restricted data regime: one year of initializations (2018) for training, 2019 for tuning and a held-out 2020 for evaluation. The training windows of the constituent ML models do not overlap the evaluation year, and the IFS baselines were generated operationally, so no leakage arises at either the input or the verification end. The near-universal lead obtained under this restriction therefore indicates that a single year of paired forecasts suffices for TaiJi to be effective; a longer training window would be expected to add skill at modest additional cost.

The ability to learn from a short training window is also valuable operationally. The system being combined is not stationary: centers upgrade their operational configurations, ML models are re-released, and the large-scale circulation varies interannually, so the relative strengths on which the combiner depends drift with time. A framework that generalizes from one year of paired forecasts can be refitted as the pool and the climate evolve, tracking the prevailing regime instead of averaging over decades of regimes that no longer apply.

## Outlook

Three directions follow from this work. The first is to enlarge the constituent pool and identify where the approach saturates. The combiner is agnostic to the identity, architecture and training pipeline of its inputs, so members such as FengWu, NeuralGCM, Aurora,

GenCast and the operational suites of additional centers can be admitted without architectural change. This has practical as well as scientific value: as competing guidance products proliferate, the burden of reconciling them falls on human forecasters, who must weigh several disagreeing deterministic and ensemble products under operational time pressure. An objective combiner calibrated on past skill can make part of this judgment reproducible and verifiable.

The second is to develop downstream applications on the combined field. Because the combiner already outperforms every member of the available pool, second-stage fine-tuning toward a specific objective inherits that skill at negligible additional cost. Heavy-precipitation forecasting, demonstrated above, is the most immediately valuable case, since precipitation extremes are where current guidance is weakest[50,51,64–66] and where constituent disagreement is largest.

The third is to refine the combination rule itself. Our ablation experiments indicate one concrete step: correcting member biases before combination reduces RMSE relative to combining raw members, suggesting that debiasing and weighting may be better treated as separate stages, provided that the accompanying loss in ACC can be avoided. Beyond this, rules conditioned explicitly on synoptic-scale flow indices and teleconnection states, and adaptive member-selection policies learned by reinforcement learning rather than fixed by regression, appear the most promising avenues.

Finally, TaiJi is not intended to replace large forecast models, and our results are not an argument against continued investment in them. Our collective account of atmospheric evolution remains incomplete, and each new architecture contributes an inductive bias that no downstream combination can create. TaiJi provides an inexpensive framework for turning each such contribution into forecast skill where it is strongest; the more capable and diverse the available models become, the more a combiner can gain from them.

# Methods

## Problem statement

We formulate global weather forecast combination as a spatiotemporal regression problem. The global verification grid is discretized at 1.5° resolution into 121 × 240 = 29,040 grid points, and forecasts are evaluated at the 12-hour cadence from 12 to 240 h. The constituent pool comprises five forecast systems: three machine-learning models (Pangu-Weather, GraphCast and FuXi) and two ECMWF operational products (IFS-HRES and the IFS-ENS mean). The IFS systems define the operational reference for global medium-range prediction, and each of the three machine-learning models has been published in the peer-reviewed literature, with archived forecasts for the evaluation period distributed through WeatherBench 2[24]. A dedicated combiner is trained for each target variable, and each combiner is optimized jointly over all 20 lead times (12–240 h), so that a single forward pass yields the weights and residuals for every lead time. Within each combiner, the atmospheric state at analysis time is encoded into a high-dimensional latent representation that conditions both the per-model weight fields and an additive residual field, and the network parameters are optimized iteratively to minimize the error of the combined forecast.

All constituents, including IFS-HRES and IFS-ENS, are verified against ERA5 throughout this study. This departs from the WeatherBench 2 convention, under which operational IFS forecasts are verified against the IFS's own analysis[24]. The departure is required by the combination task itself: TaiJi is fitted to ERA5, and its output blends all constituents, so the combined forecast and each of its members must be scored against a single reference for their errors to be comparable. As a consequence, the errors reported for the IFS systems include the discrepancy between the operational analysis and ERA5, which is largest at the shortest lead times and diminishes as forecast error grows[24]. Comparisons involving IFS-HRES and IFS-ENS in the main text should therefore be read as measuring agreement with ERA5.

The TaiJi forecast combination at each grid point and lead time is defined as

$$\hat{y}(x,\tau) = \sum_{i=1}^{M} w_i(x,\tau;\theta)\cdot f_i(x,\tau) + r(x,\tau;\theta) \tag{2}$$

where the M weight fields (M = 5 for most variables; M = 4 for Q700 and 6-h precipitation, for which one constituent does not provide the field) and the residual correction are jointly produced by a single neural network, TaiJi, whose parameters are learned end-to-end on the forecast objective. Two properties distinguish Equation (2) from classical post-processing schemes. First, the weights are functions of the instantaneous atmospheric state and are resolved at every grid point, rather than being fixed scalars or slowly varying climatologies. Second, the additive residual supplies a correction that no weighted blend of the constituents can express, thereby mitigating the systematic error that is common to the ensemble.

## Theoretical analysis: a no-harm property of pointwise affine combination

The combination rule of Equation (2) admits a simple guarantee. Fix a grid point x and a lead time $\tau$, let y denote the verifying value, and let $s = (x_0, f)$ denote the information available to the combiner, namely the analyzed atmospheric state $x_0$ and the constituent forecasts f. For a given loss $\ell$, write $R_i = E[\ell(f_i, y)]$ for the expected loss of constituent i and $R^*$ for the smallest expected loss attainable by combiners of the form of Equation (2), in which the weights and the residual may depend on s. Setting $w_i = 1$, $w_j = 0$ for $j \neq i$ and $r = 0$ reproduces constituent i exactly, so every constituent belongs to this family and $R^* \leq \min_i R_i$: the optimal pointwise affine combiner is never worse than the best constituent at that grid point and lead time. Because the weights may vary with s, the family also contains the state-dependent selection rule that chooses, for each s, the member with the lowest conditional expected loss, which yields the sharper bound $R^* \leq E_s[\min_i E(\ell(f_i, y) \mid s)] \leq \min_i R_i$. The second inequality is strict whenever the identity of the best member changes with the atmospheric state, which is the situation documented in Figure 1. The same argument quantifies the value of the background field. A combiner whose weights depend only on the forecasts f is a special case of one that depends on $s = (x_0, f)$, so its best attainable loss is no smaller than $R^*$, and it is strictly larger whenever the analyzed state $x_0$ carries information about the member errors beyond that contained in f.

Two qualifications apply. First, the bound concerns the optimum of the combiner family rather than the trained network: a model fitted on finite data may fall short of that optimum, so the property motivates the design but does not by itself certify the fitted combiner, whose performance must be established empirically (Figures 1–3). Second, the bound holds for the loss being minimized. TaiJi is trained with a latitude-weighted $L_1$ loss, so the guarantee applies formally to the mean absolute error; the advantage in RMSE and ACC reported above is an empirical finding rather than a direct consequence of the bound. Because the bound holds at every grid point and lead time, it also holds for any area-weighted average of the per-point losses.

# Data and preprocessing

## Input channels

**ERA5 atmospheric state.** Eighty-four ERA5[48] reanalysis variables supply the atmospheric conditioning context, together with a static land–sea mask. The pressure-level variables comprise temperature, humidity, geopotential, wind speed and the zonal and meridional wind components at 13 standard pressure levels (50, 100, 150, 200, 250, 300, 400, 500, 600, 700, 850, 925 and 1000 hPa); the single-level variables comprise 2 m temperature, sea surface temperature, 10 m zonal and meridional wind, mean sea-level pressure, and the 6-hour accumulated precipitation. For each forecast initialization, the three most recent analysis times, at the initialization time and six and twelve hours before it, are stacked along the channel dimension for all 84 variables and concatenated with the land–sea mask. This yields an atmospheric-context tensor of 253 channels on the full 121 × 240 grid. This tensor is processed by the ERA5 encoder; stacking three analysis times gives the network access to the recent tendency of the atmospheric state, which helps it to identify evolving synoptic regimes.

**Constituent forecasts.** At each of the 20 evaluated lead times, the M constituent forecasts are augmented with two ensemble statistics, the multi-model mean and the multi-model spread, and M corresponding anomaly fields, each member minus the multi-model mean, giving a block of $2M + 2$ channels per lead time. The anomaly and spread channels are algebraic functions of the raw member fields and are supplied explicitly so that the encoder is not required to recover the geometry of inter-model disagreement from the raw forecasts alone. The blocks for all 20 lead times are concatenated along the channel dimension, yielding a constituent-forecast tensor of $20(2M + 2)$ channels on the full grid. Presenting the complete lead-time stack, rather than a single target horizon, exposes the magnitude and spatial organization of inter-model disagreement along the entire forecast trajectory, which is informative both for weight inference and for residual estimation.

**Normalization, training, validation and test splits.** All ERA5 and forecast variables are normalized to zero mean and unit variance using statistics estimated over the 2005–2017 climatological window, and any non-finite values arising from missing or corrupted source records are replaced with zero following normalization. Training uses forecast initializations from 2018, validation uses 2019, and testing uses 2020, with the temporal order across the three splits strictly enforced; all hyperparameters, including the learning-rate schedule and early stopping, are determined solely on the validation set, precluding any leakage of information from the test period. Consistent with the WeatherBench 2 protocol, all evaluation metrics are computed on the latitude-weighted global grid at 1.5° resolution, with no further bias correction applied at evaluation time.

## TaiJi Combiner architecture

TaiJi is a dual-branch encoder–decoder neural network with an asymmetric design, comprising four interacting modules: an ERA5 encoder that receives the ERA5 background field and extracts a hierarchical representation of the atmospheric state; a forecast encoder that receives the forecasts of all constituent models at all 20 lead times and compresses them into a joint representation; a Feature-wise Linear Modulation (FiLM) interface through which the atmospheric state shapes the representation of the constituent forecasts; and a decoder with skip connections that reconstructs, in a single forward pass, full-resolution combination weights and residual corrections for all 20 lead times, so that the output weight tensor has one slice per lead time.

Although the two bottleneck representations are also concatenated before decoding, the branches are coupled primarily through the FiLM interface, so that the ERA5 encoder controls how the forecast features are transformed while the forecast encoder determines their content (see Feature-wise Linear Modulation below). The architecture is presented schematically in Figure 11, and described component-wise below.

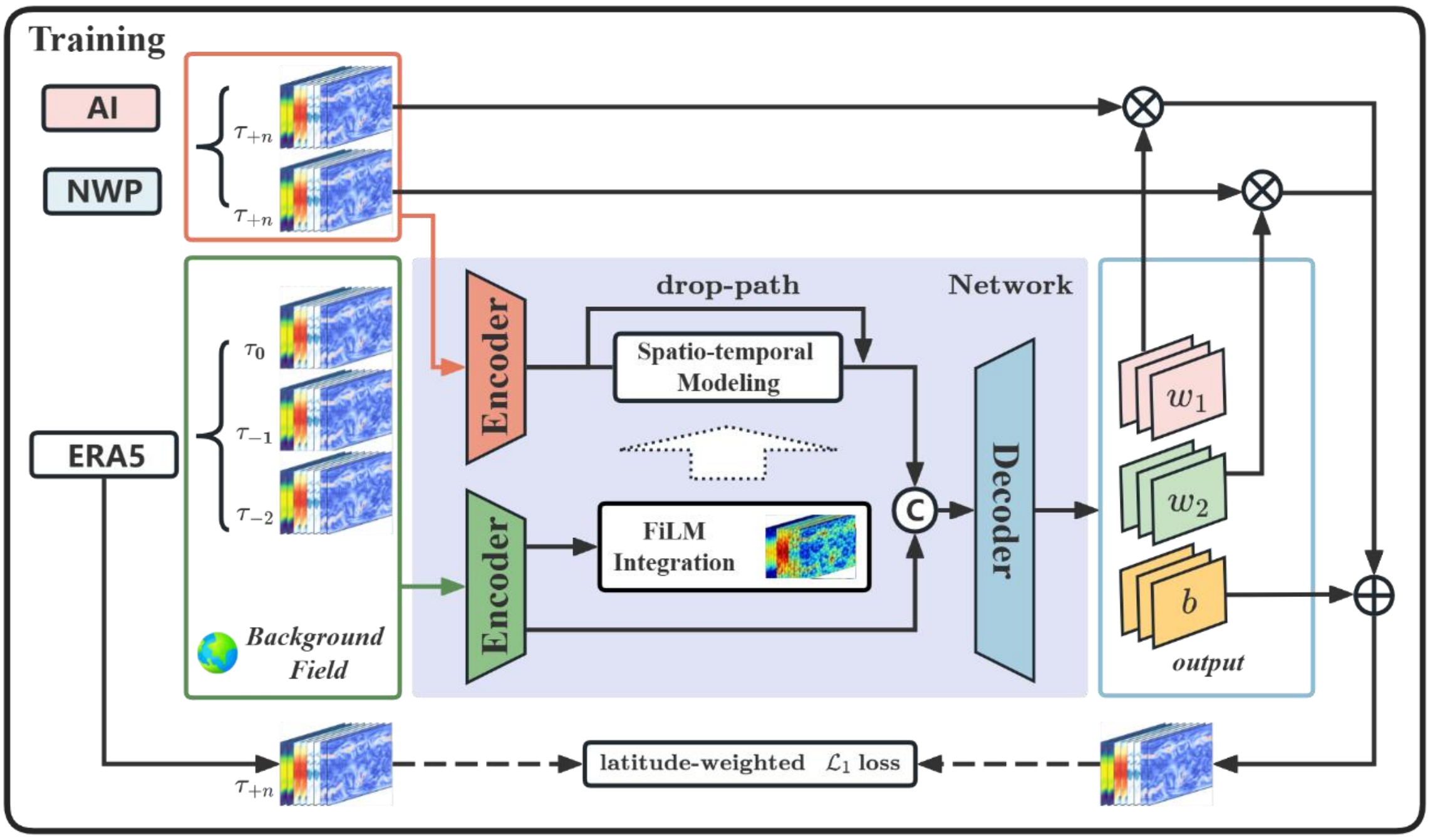

**Figure 11.** Architecture and workflow of TaiJi. The ERA5 background field at the three most recent analysis times ($\tau_0$, $\tau_{-1}$, $\tau_{-2}$) and the AI and NWP constituent forecasts at all lead times ($\tau_{+n}$) are processed by separate encoders. The atmospheric-state representation modulates the forecast representation through FiLM; the two representations are then concatenated and decoded into pixel-wise weight fields ($w_1$, $w_2$, …) and a residual field (b in the figure; r in Equation (2)). The weighted sum of the constituent forecasts plus the residual is trained against ERA5 with the latitude-weighted $L_1$ loss.

### ERA5 encoder

The ERA5 branch is a four-stage 2D convolutional encoder. The 253-channel atmospheric-context tensor is first passed through a stem convolution of output dimension 96, followed by three encoder stages, each consisting of two residual blocks[67] in pre-activation form at the current resolution and a stride-2 down-sampling step. The encoder produces three intermediate-resolution feature maps,

$$s_1 \in \mathbb{R}^{96\times H\times W},\ \ s_2 \in \mathbb{R}^{192\times (H/2)\times (W/2)},\ \ s_3 \in \mathbb{R}^{384\times (H/4)\times (W/4)} \tag{3}$$

and a bottleneck representation,

$$z_{\mathrm{ERA5}} \in \mathbb{R}^{768\times (H/8)\times (W/8)} \tag{4}$$

where H and W denote the dimensions of the verification grid. Each residual block employs the Group Normalization–ReLU–Conv (3 × 3) pattern twice, with a skip connection between the input and the output. The group count is set to eight where possible and reduced accordingly when the channel count is not divisible by eight. The three intermediate feature maps serve as skip connections to the corresponding stages of the decoder, providing the fine-scale spatial information needed to reconstruct grid-point weight fields.

**Constituent-forecast encoder**

The constituent-forecast branch processes the multi-lead-time forecast tensor through a stem convolution of output dimension 96, group-normalized with a single group, which is equivalent to layer normalization per spatial position, followed by three stride-2 down-sampling blocks, producing

$$u \in \mathbb{R}^{384 \times (H/8) \times (W/8)} \tag{5}$$

which matches the spatial footprint of the ERA5 bottleneck. Because all lead times enter as channels of a single tensor, contrasts between horizons are available to the encoder in its first convolutional layer, and the resulting representation is shared by every output horizon of the decoder. For the Q700 and precipitation tasks, where M = 4, the stem accepts a correspondingly narrower input tensor; the remainder of the architecture is unchanged.

**Feature-wise Linear Modulation**

Adaptive combination requires that the mapping from constituent forecasts to combination weights itself vary with the atmospheric state: the error structure that determines the appropriate weighting is regime-dependent, so the atmospheric state acts less as an additional predictor than as a variable that modifies the regression itself. Concatenating the ERA5 context to the forecast channels alone would express this relation only implicitly, requiring a single convolutional pathway to disentangle conditioning information from predictive content while representing both through the same additive parameterization.

We therefore couple the two branches through Feature-wise Linear Modulation, in which the atmospheric representation acts multiplicatively upon the forecast representation rather than alongside it. A global average pool collapses the ERA5 bottleneck to a 768-dimensional code, which is linearly projected to a pair of 384-dimensional scale and shift vectors; the forecast representation is then modulated channel-wise,

$$\tilde{u} = \gamma \odot u + \beta, \quad (\gamma, \beta) = Wh + b \tag{6}$$

where ⊙ denotes channel-wise multiplication, h is the pooled 768-dimensional atmospheric code, and W and b are the weight matrix and bias of the single affine projection that produces the scale vector γ and the shift vector β. Multiplicative interactions of this form generalize concatenation-based conditioning and can represent conditional dependencies that additive conditioning cannot[49,68]. Functionally, the arrangement is a lightweight hypernetwork[69]: the atmospheric state selects a member of a family of forecast-to-weight mappings, rather than contributing an additive term to a single fixed mapping.

The design is also economical. All conditioning information passes through a low-dimensional modulation code, so the capacity that the network devotes to regime dependence is governed by the width of that code rather than by the parameter count of the pathway it modulates. The projection is initialized so that the scale vector is unity and the shift vector is zero, which corresponds to the identity transformation on the forecast features; the network therefore departs from an un-modulated, state-independent combination only to the extent that the training data support it.

**Decoder**

The decoder receives the concatenated bottleneck representation, 1152 channels at one-eighth resolution, and reconstructs a full-resolution output through three up-sampling stages, each consisting of bilinear up-sampling, a residual block, and a Conv-GroupNorm-ReLU module. At every stage the up-sampled features are concatenated with the corresponding ERA5 skip connection along the channel dimension. A final 1 × 1 convolution maps the 96-channel decoder output to 20(M + 1) channels, providing at each lead time a pixel-wise weight field and a residual correction field,

$$w(\cdot,\tau) \in \mathbb{R}^{M\times H\times W},\ \ r(\cdot,\tau) \in \mathbb{R}^{H\times W} \tag{7}$$

with H and W as above. The weight channels are initialized with bias 1/M and weight standard deviation 0.01, so that the initial prediction is approximately the multi-model mean and the gradient signal flows through a well-conditioned region of the loss surface from the first optimization step.

**Composition**

The final TaiJi prediction at each grid point and lead time is the point-wise weighted combination of the constituent forecasts followed by the additive correction,

$$\hat{y}(x,\tau) = \sum_{i=1}^{M} w_i(x,\tau)\cdot f_i(x,\tau) + r(x,\tau) \tag{8}$$

which is the discretized implementation of Equation (2).

## Per-variable specialization

A separate TaiJi combiner is fitted for each of the eight target variables, yielding eight specialized combiners in total, each of which produces the complete 20-lead-time output for its variable. The rationale for this decomposition, which removes the conflict between loss surfaces of variables with very different units and dynamical ranges, is given in the Discussion. Lead times are retained within a single task because the error statistics of a given variable evolve continuously with horizon and share common physical units, so joint estimation across horizons acts as a source of regularization rather than of interference. The eight tasks are mutually independent and are trained in parallel. The precipitation combiner used in the heavy-precipitation evaluation is trained in the same way on 6-h accumulated precipitation before the second-stage fine-tuning described in Results.

## Loss function

Training uses a latitude-weighted $L_1$ objective that accounts for the unequal areal representation of grid cells in a regular latitude–longitude discretization. For grid row k at latitude φ, the cosine-of-latitude weights are

$$\omega_k = \frac{\cos\varphi_k}{\frac{1}{n_\varphi}\sum_{k'}\cos\varphi_{k'}}, \quad \sum_k \omega_k = n_\varphi \tag{9}$$

where the denominator is the mean cosine of latitude over the grid, so that the weights sum to the number of latitude rows. For a training sample, the loss aggregated over all lead times is

$$\mathcal{L}_n(\theta) = \frac{1}{NHW}\sum_{\tau\in T}\sum_{k=1}^{H}\sum_{l=1}^{W}\omega_k \left| y_n(x_{k,l},\tau) - \hat{y}_n(x_{k,l},\tau;\theta)\right| \tag{10}$$

and the empirical objective is the mean of this quantity over the training initializations. Latitude weighting places the training objective in the same areal measure as the latitude-weighted verification metrics[24,70], preventing the systematic mismatch that arises when uniformly weighted losses are used to optimize area-weighted diagnostics. The $L_1$ form is robust to the heavy-tailed error distributions characteristic of mesoscale weather features, where a small number of large-magnitude residuals would otherwise dominate the gradient. For strongly skewed variables such as precipitation, however, its optimum is the conditional median, which underestimates heavy events; this is addressed by the second-stage fine-tuning described in Results.

## Training protocol

Each of the eight tasks is optimized independently with the AdamW algorithm[71,72] at an initial learning rate of $5 \times 10^{-4}$ and a weight decay of $10^{-2}$; global gradient norms are clipped at unity to suppress occasional large updates that arise when the network encounters initializations far outside the climatological envelope. A ReduceLROnPlateau learning-rate schedule reduces the learning rate by a factor of 0.5 after three epochs of no improvement in the validation loss, with a lower bound of $10^{-7}$.

Training proceeds for up to 200 epochs with early stopping triggered after 15 epochs of no improvement in latitude-weighted RMSE on the 2019 validation set; the checkpoint corresponding to the best validation RMSE is retained as the final model. Batch size is fixed at 32. Apart from the second-stage precipitation fine-tuning (Equation (1)), only the loss of Equation (10) is back-propagated. The implementation is in PyTorch, and training is conducted on two consumer-grade NVIDIA RTX 4090 GPUs with the eight tasks scheduled across the two devices. The total training cost of the complete suite is about 16 GPU-hours (about 8 h on two RTX 4090 GPUs for all eight variables), at least an order of magnitude below that reported for any of the ML constituents; for example, pre-training FuXi alone is reported to take about 30 h on eight A100 GPUs, followed by about two days of fine-tuning on eight A100 GPUs for each cascaded model[18].

## Data availability

All datasets used in this study are publicly available. Both the atmospheric-state context supplied to TaiJi and the ground-truth verification target are drawn from the ERA5 reanalysis, distributed by the European Centre for Medium-Range Weather Forecasts (ECMWF) through the Copernicus Climate Change Service (C3S) Climate Data Store (https://cds.climate.copernicus.eu/).

The deterministic and ensemble forecasts of the IFS (IFS-HRES and IFS-ENS) and the archived forecasts of the three machine-learning constituents (Pangu-Weather, GraphCast and FuXi) are obtained at 1.5° resolution from the WeatherBench 2 public archive (https://weatherbench2.readthedocs.io/), which provides the canonical pre-processed forecasts used in our evaluation.

Population data used to define the densely populated areas in the heavy-precipitation evaluation are obtained from the Gridded Population of the World, Version 4, Revision 11 (GPWv4.11), distributed by the NASA Socioeconomic Data and Applications Center (SEDAC, https://doi.org/10.7927/H4JW8BX5).

## Code availability

The training and inference code for TaiJi will be made publicly available upon acceptance of this manuscript.

## Acknowledgments

This study is funded by the National Natural Science Foundation of China. We thank the WeatherBench 2 team for making the forecast archive and evaluation framework used in this study publicly available.